\documentclass[final,5p,times,numbers]{elsarticle}
\let\oldbibliography\thebibliography
\renewcommand{\thebibliography}[1]{%
  \oldbibliography{#1}%
  \setlength{\itemsep}{2pt}%
}
\usepackage{enumitem}
\usepackage{amsmath}
\usepackage{lipsum}
\journal{Nuclear Physics B}
\usepackage{graphicx}
\usepackage{url}
\usepackage{hyperref}
\usepackage{graphicx}
\usepackage[utf8]{inputenc}
\usepackage{color}
\usepackage{epsfig}
\usepackage{float}
\newcommand{\BE}{\begin{equation}}
\newcommand{\EE}{\end{equation}}

\newcommand{\oneloops}{\mbox{\scriptsize 1-loop}}
\begin{document}
\begin{frontmatter}
\title{Reply to ``There is no 690 GeV resonance''}
\author[1]{Maurizio Consoli}
\ead{maurizio.consoli@ct.infn.it}
\affiliation[1]{organization=
{Istituto Nazionale di Fisica Nucleare, Sezione di Catania},
            country={Italy}}
\author[2]{Leonardo Cosmai}
\affiliation[2]{organization=
{Istituto Nazionale di Fisica Nucleare, Sezione di Bari},
            country={Italy}}
\ead{leonardo.cosmai@ba.infn.it}
\author[3]{Fabrizio Fabbri}
\affiliation[3]{organization=
{Istituto Nazionale di Fisica Nucleare, Sezione di Bologna},
            country={Italy}}
\ead{Fabrizio.Fabbri@bo.infn.it}
\author[4]{George Rupp\corref{cor4}}
\affiliation[4]{organization=
{Centre for Theoretical Particle Physics, IST, University of Lisbon},
            country={Portugal}}
\ead{george@tecnico.ulisboa.pt}
\cortext[cor4]{Corresponding author}
\begin{abstract}
A recent paper has criticised the idea that, beside the resonance of mass
$m_h= 125$~GeV, the Higgs field might exhibit a relatively narrow, second
resonance with a mass $M_H \sim 690$~GeV. Without considering the evidence
we have provided, the criticism also concerned our claim that experimental
signals for this new resonance might already be seen in some LHC data. Since
our extensive work is covered by several papers, we will summarise here the
whole issue, namely: i) the theoretical motivations for a two-mass structure
in cutoff $\Phi^4$ theory; ii) the checks from lattice simulations and the
prediction $(M_H)^{\rm Theor} =  690\,(30)$~GeV; iii) the present experimental
indications of a new, relatively narrow resonance in the expected mass range.
This compact presentation will thus give the elements to objectively judge on
a relevant question of present-day particle physics.
\end{abstract}


\end{frontmatter}

\section{Introduction}

The motivation of the present article is a recent paper by J.~M.~Cline
\cite{cline}, in which he criticises the idea that, besides the known
resonance with mass $m_h= 125$~GeV, there might be a second resonance of
the Higgs field with a theoretical mass $(M_H)^{\rm Theor} = 690\,(30)$~GeV.
He also disputes our claim that in some LHC data there may be indications
supporting the existence of such a second resonance. Since our work, mainly
spread over Refs.~\cite{Cosmai2020,EPJC,EPL}, concerns many theoretical and
phenomenological aspects, we will present here a summary of the main ideas.
We believe that this more compact presentation will provide the elements to
objectively judge on an important question of field theory and
particle physics. 

The experimental observation of the narrow scalar resonance with mass
$m_h= 125$~GeV and the consistency of its phenomenology with the theoretical
expectations for the Higgs boson have confirmed Spontaneous Symmetry Breaking
(SSB), through an underlying $\Phi^4$ theory, as the essential ingredient
determining the Standard Model (SM) vacuum. However, the generally accepted
``triviality''\footnote
{There have been in the past attempts to avoid ``triviality''. Some of these
efforts are based on Symanzik's original observation \cite{kurt} that a
$\Phi^4$ theory with negative bare coupling would be asymptotically free. Due
to the instability of the corresponding potential, the negative-coupling
theory was never taken seriously; see nonetheless the recent work by
Romatschke \cite{Romatschke} in the somewhat different context of a
$\Phi^4$ with ${\mathcal O}(N)$ symmetry. Another argument, concerning the
``triviality'' vs.\ asymptotic freedom ambiguity, came from considering
$\Phi^4$ in the context of Wilson's $\epsilon$ expansion \cite{huang}. Due
to the different nature of the two fixed points in $d=4+\epsilon$, for
$\epsilon \to \pm 0 $, it is not surprising that the single fixed point
into which they collapse has both infrared and ultraviolet aspects. In the
context of the effective potential, it is as if there were two separate
$\Phi^4$ theories \cite{paul1987} inhabiting $d=4 \pm \epsilon$ dimensions.
Here, as in Refs.~\cite{Cosmai2020,EPJC,EPL}, the currently accepted
``triviality'' scenario is assumed.}
of $\Phi^4$ theories in four space-time dimensions indicates a trivially
free continuum limit. Therefore, even though SSB through the Higgs field
works remarkably well, this would represent, in the end, an effective
description. 

But in a technical sense, a ``trivial'' theory exhibits a Gaussian
distribution of Green's functions,
including a first moment $\langle \Phi\rangle \neq 0$ of this distribution.
Therefore, to check the present view of a simple massive theory with small
perturbative corrections, it is natural to consider those Gaussian
approximations to the effective potential and effective action in which
the fluctuation fields are governed by a quadratic Hamiltonian with optimised
parameters. This is even more true when considering that, in this type of
approximations and differently from standard perturbation theory, SSB in
$\Phi^4$ is described as a weak first-order (or quasi-first-order) phase
transition, consistently with the indications of lattice simulations
\cite{lundow2009critical,Lundow:2010en,akiyama2019phase,giapponesi2}. 

It is this type of Gaussian calculations, for both single-com\-ponent and
continuous-symmetry ${\mathcal O}(N)\,\Phi^4$ theories, that lead to a
Euclidean propagator $G(p)$ depending on two mass parameters, viz.\ $m_h$
and $M_H$. These are associated with the $p \to 0$ and large-$p^2$ limits
of $G(p)$, respectively, and scale distinctly with the logarithm
$L \sim \ln (\Lambda/M_H)$ of the Euclidean cutoff $\Lambda$, that is,
$m_h\sim L^{-1/2} M_H$. As such, the two masses play different roles in the
description of SSB, because $m_h$ defines the quadratic shape of the potential
at its minimum, whereas the fourth power of $M_H$ determines the size of the
zero-point energy (ZPE) $\sim -M^4_H L$ as well as the potential depth
${\Delta V}\sim -M^4_H$. We emphasise that the different scaling of the two
masses is crucial for the continuum theory to have only one scale. But in the
cutoff theory we have both. This is why the physical picture is substantially
different from standard perturbation theory, even though there is no
contradiction with  the ``triviality'' of the theory.  

Lattice simulations of the scalar propagator have been carried out
\cite{Cosmai2020} to check the existence of two regions in Euclidean momentum
space where $G^{-1}(p)\sim (p^2 + m^2_h)$ and  $G^{-1}(p)\sim (p^2 + M^2_H)$.
These simulations, while confirming the expected asymmetric scaling
$m_h\sim L^{-1/2} M_H$ also lead to a numerical prediction for $M_H$. Moreover,
$m_h$ indeed vanishes as $L^{-1/2}$ in units of the Fermi scale
$v\sim $246 GeV. Therefore, the larger mass $M_H$ would remain finite through
a simple proportionality relation $M_H= Kv$. With the lattice determination
$K_{\rm latt}\sim 2.81(12)$, a theoretical prediction for $M_H$ was obtained,
viz.\ \cite{Cosmai2020}
\BE
(M_H)^{\rm Theor} =  690\,(30) \; {\rm GeV } \; .
\label{prediction}
\EE
These lattice simulations were performed in a single-component $\Phi^4$ theory,
so one may wonder what the relevance is for the physical Higgs field with its
${\mathcal O}(4)$ symmetry. However, $m_h$ and $M_H$ refer to different
geometric properties of the effective potential, which is rotationally
invariant. Therefore, the relationship between these two mass parameters
should not depend on the number of field components. A more quantitative
argument is that, as $\Lambda$ gets smaller, $m_h$ increases in units of $M_H$,
reaching its natural upper limit when $\Lambda$ becomes as small as possible:
\BE
(m_h)^{\rm max} \sim (M_H)^{\rm Theor} = 690\,(30)  \; {\rm GeV } \; .
\label{ourbound}
\EE    
This definite prediction can thus be compared to the known upper bounds on
$m_h$, obtained many years ago from lattice simulations of the
${\mathcal O}(4)$ theory. In this case, a very good agreement is found with
Lang's \cite{lang} and Heller's \cite{heller} estimates, viz.\
$(m_h)^{\rm max}=670\,(80)$~GeV and $(m_h)^{\rm max}=710\,(60)$~GeV,
respectively, depending on slightly different assumptions about the
magnitude of the minimum ultraviolet cutoff. Actually, by combining
these two estimates from the ${\mathcal O}(4)$ theory, the result
\BE
(m_h)^{\rm max} \sim 690\,(50)  \; {\rm GeV } \;\;\;\;
({\rm Refs.}\,\mbox{\cite{lang,heller}} \; {\rm combined})
\label{langheller}
\EE 
practically coincides with our expectation from the one-compo\-nent theory
in Eq.~(\ref{ourbound}), thus yielding a consistency check of the whole
two-mass scheme. At the same time, since in the real world $m_h\!=\!125$~GeV,
the substantial difference between $m_h$ and $M_H$ would mean that the 
$\Phi^4$-sector cutoff is extremely large. 

After these foundational aspects, to be further discussed in Sec.~2, we will
briefly recall in Sec.~3 the phenomenological profile of the new resonance
as a relatively narrow resonance and then the experimental evidence
supporting the existence of such a new resonance in the expected mass range. 
However, to fully appreciate some of this evidence, one should definitely go
beyond the narrow-width approximation, where interference effects become
irrelevant and, as for the 125 GeV Higgs resonance, one simply looks for a
sharply localised excess above a smooth background. Instead, when the decay
width is comparable to, or larger than, the experimental resolution, the
interference effect, changing sign across the resonance peak, can become
crucial. In particular, as in our case where the resonance peak cross section
$\sigma_P$ is comparable to, or smaller than, the background $\sigma_B$, the
interference term, proportional to $\pm  \sqrt { \sigma_P\sigma_B  } $, could
determine a characteristic ``peak-dip'' sequence. 

\section{The two-mass structure}

We will recall here the basic steps that lead to the $(m_h,M_H)$ structure of
the scalar propagator in a single-component $\Phi^4$ theory. For the case of
a scalar field with an ${\mathcal O}(N)$ symmetry, we refer to Subsec.~2.5
of Ref.~\cite{EPJC}. While the calculations are considerably more involved, the main aspects remain the same. 

\subsection{The Gaussian Effective Potential}

Let us start for definiteness from the usual picture of SSB as a second-order
phase transition with a classical potential 
\BE
U(\phi) \; = \; \frac{1}{2} m^2_B \phi^2 + \frac{\lambda}{4!}\phi^4 \; .
\label{classical}
\EE
Denoting its non-zero minima by $\phi=\pm v$, where $m^2_B+(\lambda v^2)/6=0$,
the quadratic shape of the potential at these minima is then defined as
\BE
\left[\frac{d^2 U(\phi)}{d \phi^2}\right]_{\phi=\pm v} \equiv \; m^2_h
\; = \; \frac{\lambda } {3} v^2 \; .
\EE
These relations are assumed to remain valid to all orders in perturbation
theory and used to define the Fermi scale $v\sim$ 246 GeV and the Higgs boson
mass $m_h=$ 125 GeV. In that sense, this pair of fundamental quantities are
inextricably linked to each other by also constraining any non-perturbative
calculation of the effective potential $V_{\rm eff}(\phi)$ to yield 
\BE
\left[\frac{d^2 V_{\rm eff}(\phi)}{d\phi^2}\right]_{\phi=\pm v} \equiv \;
m^2_h \; = \; \frac{\lambda} {3} v^2 \; . 
\label{general}
\EE
In pure $\Phi^4$, as a consequence of ``triviality'', when
$\lambda=\lambda(\mu)$ vanishes proportionally to
$ L^{ -1}=[\ln (\Lambda/\mu)]^{-1}$, one would thus deduce
$(m^2_h/v^2) \sim L^{ -1}\to 0$ in the $\Lambda \to \infty$ limit.  

After this premise, let us now consider the description of SSB in the
Gaussian approximation to the effective potential $V_{G}(\phi)$; see
Ref.~\cite{barnes,gaussian,ciancitto}. Here, differently from the potential
in Eq.~(\ref{classical}), SSB is a weak first-order (or quasi-first-order)
phase transition consistently with the indications of lattice simulations
\cite{lundow2009critical,Lundow:2010en,akiyama2019phase,giapponesi2}. In fact,
the classically scale-invariant theory, defined by
$[d^2 V_{G}(\phi) / d \phi^2]_{\phi=0 }= 0$, is found in the broken-symmetry
phase. Therefore, the phase transition occurs when the mass squared in the
symmetric phase is infinitesimal yet still positive. This is like in the
Coleman \& Weinberg \cite {Coleman:1973jx} one-loop calculation, in which this
scale-invariant limit is defined by the condition 
\BE
0 \; = \; m^2_B + \frac{ \lambda } {2} I_0[0] \; ,
\EE
with
\BE
I_0(\Omega) \; = \; \int\frac{d^4k}{(2\pi)^4} \, \frac{1}{k^2+\Omega^2} \; .
\label{I_0}
\EE
This qualitative agreement reflects the fact that the Gaussian Effective
Potential (GEP) can also be expressed as the sum of some background potential
plus zero-point energy, with suitably redefined parameters; see
Ref.~\cite{Cosmai2020,EPJC}. 
Vice versa, this means that the one-loop potential can also admit a
non-perturba\-tive interpretation, as the prototype of those approximations
where the fluctuation field is governed by an effective quadratic Hamiltonian. 

The key ingredient of the GEP is the self-consistent solution for the 
variational mass parameter $\Omega=\Omega(\phi)$, corresponding to an
all-order resummation of the tadpole graphs. The analogous relations for a
scalar field with an ${\mathcal O}(N)$ symmetry can be found in
Ref.~\cite{alles,okopinska}. 
The expression for this mass reads
\BE
\Omega^2(\phi) \; = \; m^2_B + \frac{ \lambda \phi^2 }
{2} + \frac{ \lambda } {2} I_0[\Omega(\phi)] \; ,
\label{self}
\EE
in terms of which the minimum condition of the GEP has the form
\BE
\frac{d V_{G}(\phi)}{d \phi} \; = \; \phi \left(\Omega^2(\phi) -
\frac {\lambda \phi^2}{3}\right) \; = \;  0 \; , 
\EE
so that non-zero minima occur at those values $\phi=\pm \phi_v$ where 
\BE
M^2_H\equiv \Omega^2(\phi_v) \; = \; (\lambda \phi^2_v/3) \; .
\label{phi_v}
\EE
Furthermore, by defining
\BE
\epsilon \; = \; \frac{\lambda}{16\pi^2 }
\EE
and introducing the function $J(p,M)$ given by
\BE
J(p,M) \; = \; \lambda\int\frac{d^4k }{(2 \pi)^4} \,
\frac {1}{(k^2 + M^2)[(k+p)^2 + M^2]} \; ,
\label{JpM}
\EE 
the minimum of the effective potential can be conveniently expressed through
the following two relations:
\BE
{\Delta V}_G \; = \; V_{\rm G }(\phi_v) \; = \; -\frac {M^4_H}{128 \pi^2}
\label{ground}
\EE and 
\BE
J(0,M_H) + \epsilon \; = \; 1 \; .
\label{triv1}
\EE 
Since, in terms of the logarithm of the Euclidean cutoff
$L=\ln \frac{\Lambda}{M_H}$, one finds $J(0,M_H)= 2\epsilon L$, the relation
$\lambda \sim L^{-1}$, from Eq.~(\ref{triv1}), supports again the
``triviality'' scenario. 

\begin{figure}[hb]
\centering
\includegraphics[width=0.45\textwidth,clip]{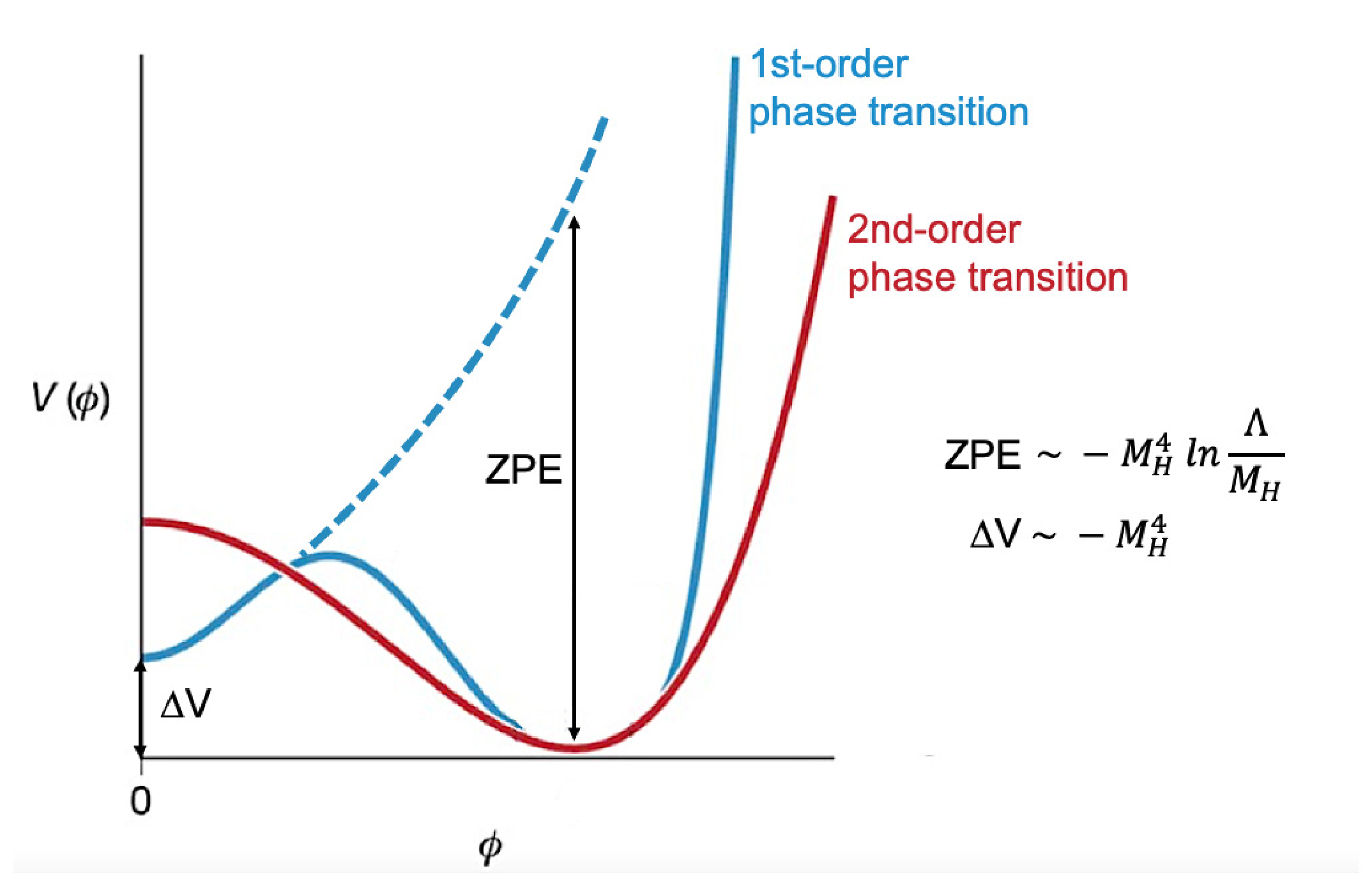}
\caption{\it The special role of the zero-point energy (ZPE) in a 
first-order scenario of SSB. Differently from a second-order
picture, it has to compensate for a tree-level potential with no
non-trivial minimum.}
\label{zpe}
\end{figure}

The most important point is that the quadratic shape of the GEP at the minima
is not $M^2_H$. Instead, one finds 
\BE 
\left[\frac{d^2 V_{G}(\phi) }{d \phi^2}\right]_{\phi=\pm \phi_v } = \; 
M^2_H\times\frac{1-J(0,M_H)}{\displaystyle1+\frac{J(0,M_H)}{2}}
\; \equiv \; m^2_h \; .
\EE 
Thus, in view of Eq.~(\ref{triv1}), we obtain
\BE
m^2_h \; \sim \;  \frac{2\epsilon}{3}M^2_H \; \sim  \; M^2_H \, L^{ -1}
\; \ll \; M^2_H \; .
\label{log}
\EE 
The same pattern is found in the Coleman-Weinberg one-loop calculation
\cite {Coleman:1973jx}. There, one is faced with a mass parameter
$M^2(\phi)=(\lambda\phi^2)/2$ whose value at the corresponding minima of the
one-loop potential, say $\phi = \pm \hat {\phi}_v$, still determines the
vacuum energy density as in Eq.~(\ref{ground}), i.e.,
\BE
\label{ground1loop}
{\Delta V}^{\oneloops} \; = \; V_{\oneloops}(\hat {\phi}_v) \; =\;
-\frac{M^4(\hat{\phi}_v)}{128 \pi^2} \; ,
\EE
which is now fixed by the condition $(3/2)[J(0,M(\hat{\phi}_v))+\epsilon]=1$.
For this reason, the quadratic shape of the one-loop potential is again much
smaller than the corresponding $M^2(\hat {\phi}_v)$:
\begin{align}
(m^2_h)^{\oneloops} =  
\left[\frac{d^2V_{\oneloops}(\phi)}{d\phi^2}\right]_{\phi=\pm{\hat\phi}_v}
& = 
M^2(\hat{\phi}_v)\,\left[1-\frac{3}{2}J(0,M(\hat{\phi}_v))\right] \nonumber \\
& \sim M^2(\hat{\phi}_v) L^{ -1} \ll M^2(\hat{\phi}_v).
\label{log1loop}
\end{align}
This close analogy between the GEP and the one-loop potential is because the
resummation of one-loop bubbles in the GEP is equivalent to the replacement
\cite{Cosmai2020,EPJC}
\BE
\lambda \to \lambda_G \; = \; \frac{\lambda}{1+\frac{J(0,M_H)}{2} } 
\; \sim \; \frac{2}{3} \lambda \; .
\EE  
To conclude this first part, a comment is due about the weak first-order vs.\
second-order ambiguity in the description of SSB. The usual point of view is
that differences in the effective potential near $\phi=0$ or at very large
$|\phi|$ are irrelevant. Any potential satisfying Eq.~(\ref{general}) would
produce the same phenomenology at the Fermi scale. By looking at
Fig.~\ref{zpe}, one can get a hint that this conclusion may be incorrect.
Indeed, the red and blue curves have the same quadratic shape at the minimum.
However, differently from the red curve, where SSB is of second order
and originates in a negative quadratic curvature at $\phi=0$, in the blue
curve SSB is of first order and induced by the zero-point energy (ZPE), which
overwhelms a tree-level potential with no non-trivial minimum. Since
the ZPE mass scale $M_H$ and the mass scale $m_h$ play different roles, one
may then expect a non-trivial difference. This expectation is confirmed by our
analysis of the GEP (and of the one-loop potential);
compare Eqs.~(\ref{ground}) and (\ref{log}). Further implications will become
clearer by looking at the scalar propagator. 

\subsection{The scalar propagator}

Let us now consider the momentum-dependent propagator. This requires to
switch from the effective potential to the effective action. For our
scope, the relevant approximation is the Gaussian Effective Action
(GEA), worked out by A.~Okopinska \cite{okopinska} in the
${\mathcal O}(N)$-symmetric case. As discussed in Subsec.~2.3 of
Ref.~\cite{EPJC}, the resulting Euclidean propagator can be obtained via
a two-step procedure: i) first reabsorbing one-loop tadpoles to all orders
into the previous Gaussian mass $\Omega(\phi)$; ii) then resumming all
(non-tadpole) one-loop bubbles with mass $\Omega(\phi)$. This second step
amounts to considering the propagator series
\begin{align}
G^{-1}(p) & = 
p^2+\Omega^2-\frac{\lambda\phi^2}{2}J(p,\Omega) 
\left\{1\!-\!\frac{J(p,\Omega)}{2}\!+\!\frac{J^2(p,\Omega)}{4}\!+\ldots\right\}
\nonumber \\
&=p^2+\Omega^2-\frac{\lambda\phi^2}{2}\frac{J(p,\Omega)}{1+ J(p,\Omega)/2}\;,
\label{propseries}
\end{align}
where $\Omega= \Omega(\phi)$ everywhere. Since at the minima
$\phi=\pm \phi_v$ of the Gaussian effective potential
$\Omega^2(\phi_v)=\lambda\phi^2_v/3=M^2_H$, we then obtain
\BE
G^{-1}(p) \; = \; p^2 +  M^2_H A(p,M_H) \; ,
\label{propagatorGEA1}
\EE
with 
\BE
A(p,M_H) \; = \; \frac{ 1-J(p,M_H) }{\displaystyle 1+\frac{J(p,M_H)}{2}} \; .
\label{Ap}
\EE
Notice that, differently from other cases where one is faced with divergent
series, here the series of one-loop bubbles is convergent for {\it all}
\/values of $p^2$, because $J(p,M_H)/2 \leq J(0,M_H)/2=(1-\epsilon)/2 < 1/2$.
Therefore, within a cutoff theory, one has to explore the entire Euclidean
range $0 \leq p^2 \leq \Lambda ^2$ or, in terms of $z =p^2/\Lambda^2$, the
whole range $0\leq z \leq 1$. This is why, by only restricting to
$0\leq p^2 \lesssim M^2_H$ as in Ref.~\cite{cline}, one ignores that the
corrections produced by $J(p,M_H)$ may become negligible, as it happens
for small enough $\epsilon$ and large enough $p^2$. Then, independently of
$A(0,M_H)$, in that region the mass would be close to $M_H$. However, a large
Euclidean $p^2$ does {\it not} \/correspond to a large
$|p^2_L|$ in the Lorentzian signature. In fact, a large Euclidean $p^2$ can
translate to a massless on-shell Lorentzian particle, that is, with $p^2_L=0$,
as well as to positive or negative values of the various kinematic invariants
(i.e., $ s > 0$  or $ t < 0$); see e.g.\ Ref.~\cite{donoghue}. Equivalently,
a very energetic proton in cosmic rays, with energy close to the GZK cutoff
\cite{GZK} $E\sim 10^{10}$~GeV, would have a minuscule
$p^2_L={\bf p}^2-E^2=-(0.938)^2$ GeV$^2$ but an enormous Euclidean
$p^2= {\bf p}^2 +p^2_4\sim 10^{20}$~GeV$^2$. For this reason, since the
combined constraint on the Euclidean cutoff, i.e.,
$\epsilon \ln (\Lambda^2/M^2_H)\sim 1$, cannot be translated into
calculations performed with the Lorentzian signature, the simplest strategy
is to first determine the two regions of Euclidean momentum where $G^{-1}(p)$
is well approximated by $\sim(p^2 + m^2_h)$ and $\sim (p^2 + M^2_H)$. Then,
from each region, one can define the Lorentzian propagator, namely
$G^{-1}_h(s)\sim(-s+m^2_h)$ and $G^{-1}_H(s)\sim(-s+M^2_H)$, respectively.
While this will not determine the function that interpolates between $G_h(s)$
and $G_H(s)$, it would still produce the picture of two quasi-particles, with
mass $m_h$ and $M_h$, which mimic the two energy branches of superfluid He-4,
density fluctuations (phonons), and vortex modes (rotons), whose hybridisation
describes the energy spectrum of the system. 

With this premise, to study the two regimes, it is convenient to give two
different expressions for the same function $J(p,M)$, namely
\BE
J(p,M) = \epsilon \left\{\ln\frac{\Lambda^2}{M^2} - \int^1_0 \! dx \,
\ln\left[1+x(1-x) \frac{p^2} { M^2} \right]\right\} 
\label{low}
\EE
and
\BE
J(p,M) = \epsilon \left\{\ln\frac{\Lambda^2}{p^2} - \int^1_0 \! dx \,
\ln\left[x(1-x) +\frac{M^2} { p^2} \right]\right\} .
\label{high}
\EE 
The deviation from the Gaussian mass can then  be expressed as
\BE
M^2(z,\epsilon) \; = \; M^2_H \, A(z,\epsilon) \; ,
\EE 
with
\BE
A(z,\epsilon) \; = \;
\frac{1-J(z,\epsilon)}{1+\frac{J(z,\epsilon)}{2}} \; .
\EE
Thus, if we take the $z\to 0$ limit for non-zero $\epsilon \ll 1$ from
Eq.~(\ref{low}), we reobtain  Eq.~(\ref{log}). Instead, to find the region of
$z$ where $M^2(z,\epsilon) \sim M^2_H$, let us choose a value $z=\bar z\neq0$
and divide the $[0,1]$ range into two regions, say I $=[0, \bar z]$ and II
$=[\bar z, 1]$. From Eq.~(\ref{high}), the function $J(\bar z,\epsilon) $ can
be expressed as
\BE
J(\bar z,\epsilon) \; = \; \epsilon\left[\ln\frac{1}{\bar z}-h(a^2)\right] \;,
\EE
with
\BE
a^2 \; = \; \frac{M^2_H} { p^2} \; = \;
\frac{1}{\bar z} \exp\left[- \frac{(1-\epsilon)}{\epsilon} \right] \; ,
\EE
\BE
h(a^2) \; = \; -2 + \ln(a^2) + \Delta \ln\frac{\Delta+1}{\Delta-1} \; ,
\EE
and
\BE
\Delta \; = \; \sqrt {1 + 4a^2} \; .
\EE
Therefore, for $\epsilon \ll 1$, where
$J(\bar z,\epsilon) \sim \epsilon (\ln (1/ \bar z) + 2)$, if we fix $\bar z$
by imposing $J(\bar z,\epsilon) \ll 1$, we find $M^2(z,\epsilon) \sim M^2_H$
in the whole region II. We emphasise that, for $\epsilon \to 0$, this remains
true for extremely small values of $\bar z$, for instance with
$\bar z \sim \epsilon^{n}$ and arbitrary $n$, the only condition being
$\exp(-1/\epsilon) \ll \bar z$. Therefore, region II will actually fill the
entire range of $z$, with the only exception of the infinitesimal region
$z\sim \exp(-1/\epsilon)$, where $M^2(z,\epsilon) \sim m^2_h$. 

\subsection{Check with lattice simulations of the propagator and the value of
$M_H$}
\begin{figure}[ht]
\centering
\includegraphics[width=0.35\textwidth,clip]{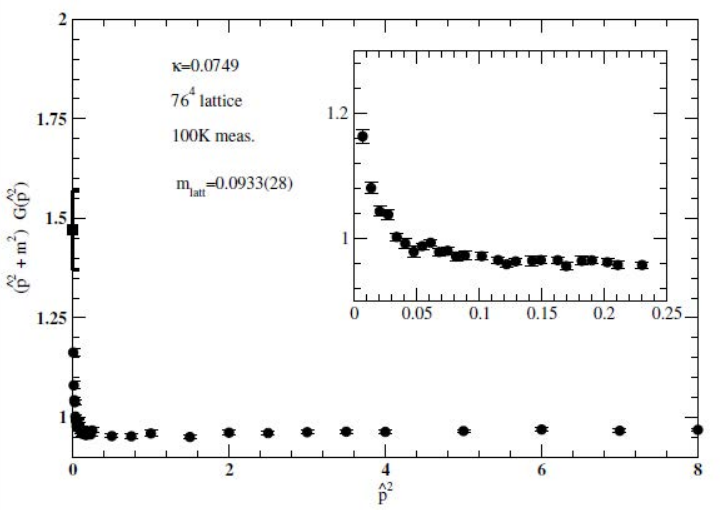}
\caption{\it The propagator data of Ref.~\cite{Cosmai2020}, rescaled by the
lattice mass $M_H \equiv m_{\rm{latt}}=0.0933\,(28)$ obtained from
fitting the data with ${\hat p}^2>0.1$. The $\hat p=0$ peak is for
$M^2_H/m^2_h= 1.47\,(9)$ with zero-momentum mass $m_h=0.0769\,(8)$.}
\label{0749}
\end{figure}
\begin{figure}[ht]
\centering
\includegraphics[width=0.35\textwidth,clip]
{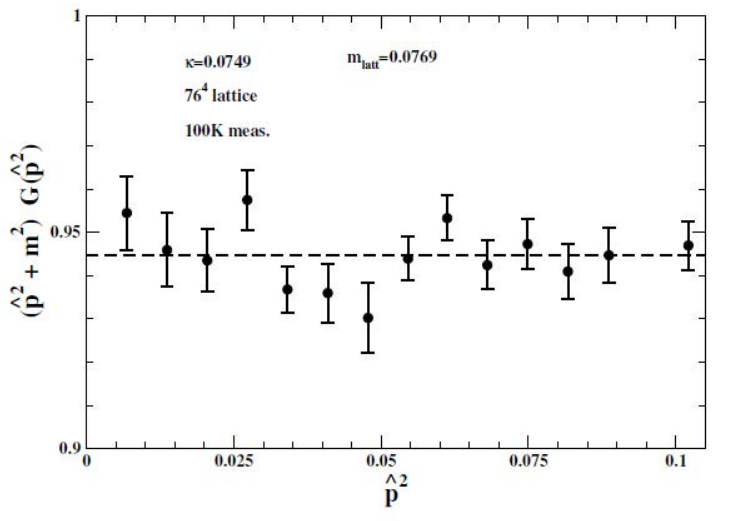}
\caption{\it The propagator data of Ref.~\cite{Cosmai2020} for
${\hat p}^2<0.1$. Here the lattice data are rescaled by $m_h=0.0769\,(8)$.}
\label{0749small}
\end{figure}
To check the existence of two regions in Euclidean space with
$G^{-1}(p)\sim(p^2+m^2_h)$ for $p\to0$ and $G^{-1}(p)\sim(p^2+M^2_H)$ at
larger $p^2$, respectively, with also $m^2_h \sim M^2_H  L^{ -1}$, lattice
simulations of the scalar propagator were carried out in
Ref.~\cite{Cosmai2020}. These were done in the Ising limit of the theory,
i.e., with a lattice coupling at the Landau pole, where the broken-symmetry
phase corresponds to values of a hopping parameter $\kappa$ larger than a
critical value $\kappa_c=0.074848(2)$ \cite{Stevenson2005}. The simulations
started at large momenta while progressively including smaller and smaller
momentum data. In terms of the squared lattice momentum ${\hat p}^2$, the
propagator data were thus first fitted to a standard two-parameter form 
$Z/({\hat p}^2+m^2_{\rm{latt}})$ and then rescaled by
$({\hat p}^2+m^2_{\rm{latt}})$, so that deviations from a straight line
become immediately visible. As noticed in Ref.~\cite{Stevenson2005} and
in contrast to the symmetric phase $\langle \Phi \rangle=0$, where a
single-mass propagator can well describe the whole range
$0 \leq p^2\leq \Lambda^2 $, in the broken-symmetry phase the fits become poor
when including the lowest momentum points; see Fig.~\ref{0749}. In that region,
one should switch to the smaller mass $m_h=(2\kappa\chi)^{-1/2}$, obtained
from the zero-momentum susceptibility $\chi$; see Fig.~\ref{0749small}.
\begin{table*}[t]
\caption{\it For various values of the $\kappa$ parameter of
the $4D$ Ising model and various lattice sizes, we report the determinations
obtained by different authors (see Ref.~\cite{Cosmai2020}) 
of the mass $M_H$ from the higher-momentum propagator data and the
zero-momentum $m_h$ extracted from the lattice susceptibility $\chi$ through
the relation $(2\kappa\chi)^{-1/2}$. For $\kappa=0.0749$, the three values of
$M_H$ correspond to high-momentum fits for ${\hat p}^2>0.1$, ${\hat p}^2>0.15$,
and ${\hat p}^2>0.2$. The lattice cutoff $\Lambda_L\sim \pi/a$ and all masses
are in units of the inverse lattice spacing $a$, so that
$L= \ln(\Lambda_L/M_H)$. In the last column, we present
the combination $(c_2)^{-1/2}= M_H \, (m_h)^{-1} \, L^{-1/2}$.}
\begin{center}
\begin{tabular}{lccccc}
$\kappa$ & {\rm lattice} & $M_H$ & $ m_h$ & $L^{-1/2}$ &
$(c_2)^{-1/2}$  \\
\hline \hline
0.07512 & $32^4$ & 0.2062\,(41) & 0.1857\,(8) & 0.606\,(2) & 0.673\,(14)\\
\hline0.0751 & $48^4$ & $\sim0.200$ & 0.1796\,(5) & $\sim0.603$ & $\sim0.671$\\
\hline0.07504 & $32^4$ & 0.1723\,(34) & 0.1507\,(7)& 0.587\,(2) & 0.671\,(14)\\
\hline0.0749 & $76^4$ & 0.0933\,(28) & 0.0769\,(8) & 0.533\,(2) & 0.647\,(22)\\
\hline0.0749 & $76^4$ & 0.096\,(4)   & 0.0769\,(8) & 0.535\,(3) & 0.668\,(31)\\
\hline0.0749 & $76^4$ & 0.100\,(6)   & 0.0769\,(8) & 0.538\,(4) & 0.700\,(42)\\
\hline
\end{tabular}
\end{center}
\label{latticetable}
\end{table*}
\normalsize
By considering various lattice sizes, it was thus possible to check
Eq.~(\ref{log}) as well as Eq.~(\ref{log1loop}), by determining a constant
$c_2$ that fixes the logarithmic slope
\BE 
M_H \times (m_h)^{-1} \times L^{-1/2} \; \sim \; (c_2)^{-1/2} \; ,
\label{lattice}
\EE 
whose fitted value turned out to be $(c_2)^{-1/2}=0.67\,(3)$; see
Table~\ref{latticetable}. 

Besides checking the predicted two-mass structure, the lattice simulations
allow a numerical determination of $M_H$. To this end, let us replace
$m^2_h=(\lambda v^2)/3$ in Eq.~(\ref{lattice}) and then use 
the leading-order relation $\lambda \sim (16 \pi^2/3)  L^{-1}$. With the
lattice value $(c_2)^{-1/2} = 0.67\,(3)$, we find that $M_H$ and $v$ are
related by a simple proportionality constant\footnote
{From $m^2_h= (\lambda v^2)/3 $ and Eqs.~(\ref{phi_v},\ref{log}), the two
fields $v$ and $\phi_v$ are related through a rescaling
$ v^2/ \phi^2_v =  m^2_h/M^2_h\sim L^{-1} $.}
$K \sim (4\pi/3)(c_2)^{-1/2}\sim 2.81\,(12)$, resulting in the numerical
prediction
\BE
(M_H)^{\rm Theor} \; = \; Kv \; = \; 690\,(30)~{\rm GeV} \; .
\EE 
The numerical value  $(c_2)^{-1/2} = 0.67\,(3)$ was extracted from lattice
simulations of a single-component $\Phi^4$ theory. But $m_h$ and $M_H$
correspond to different geometric properties of the effective potential, which
is rotationally invariant. This suggests that their relative size should not
depend on the number of field components. As anticipated in the Introduction,
a numerical check can be obtained by observing that $M_H$ places a natural
upper limit $(m_h)^{\rm max} \sim M_H$ when the cutoff $\Lambda$ becomes as
small as possible. Our prediction of $M_H$ can thus be translated into the
numerical prediction $(m_h)^{\rm max} \sim 690\,(30)$~GeV, which is in
remarkable agreement with the combined determination
$(m_h)^{\rm max} \sim 690\,(50)$~GeV obtained from lattice simulations of
the ${\mathcal O}(4)$ theory \cite{lang,heller}. At the same time, since in
the real world $m_h=125$~GeV, a second resonance with $M_H\sim 690$~GeV would
imply that the ultraviolet cutoff $\Lambda$ of the $\Phi^4$ sector is
extremely large.

\section{Comparing with experiments}

As for the expected phenomenology, we refer to Ref.~\cite{EPJC,EPL}. However,
two main points should be briefly mentioned. First, as in the original SM
perspective, due to the very large value of $M_H$ SSB would essentially be
determined by the pure scalar sector.\footnote
{The logarithmically divergent terms in the ZPE are proportional to the
fourth power of the mass, so that in units of the pure scalar term one finds
$(6 M^4_w + 3 M^4_Z)/M^4_H \lesssim 0.002$ and $12 m^4_t/M^4_H\lesssim 0.05$.}
Secondly, the pure scalar self-coupling $\lambda(\mu)$ that we have considered
in Sec.~2, which evolves towards its asymptotic Landau pole, and the
perturbative coupling $\lambda^{\rm (p)}(\mu)$, which includes the effect of
fermion and gauge fields, coincide for $\mu=v$, where both take the numerical
value $3 m^2_h/v^2\sim  0.77$. Therefore, with experiments at the Fermi scale,
their large-$\mu$ difference should remain unobservable. Confirming our picture
then requires observing the second resonance. Here, the phenomenology is not
like with a standard Higgs boson of 700~GeV. For instance, the
longitudinal-$W$s large tree-level contact coupling $\lambda_0= 3 M^2_H/v^2$
is changed to $\lambda(v)= 3 m^2_h/v^2= (m^2_h/M^2_H) \lambda_0$ after
resumming graphs to all orders. This can explicitly be shown by the
Equivalence Theorem, when understood as a non-perturbative statement, valid to
lowest non-trivial order in $g^2_{\rm gauge}$, but to all orders in the scalar
self-couplings \cite{bagger}. Analogously, the large three-linear Higgs
coupling $M^2_H/(2v)$ is reduced to $(m_h M_H)/2v$, but will be larger than
$m^2_h/(2v)$. The resulting rescaling $\kappa_\lambda=(M_H/m_h) \sim 5.5$
\cite{EPL} is nevertheless consistent with the experimental determinations
of $\kappa_\lambda$ in Refs.~\cite{atlascouplings,CMScouplings}. As a
consequence, the large conventional width $\Gamma(H \to ZZ+WW)\sim G_F M^3_H$
is suppressed by $(m_h/M_H)^2\sim 0.032$. Thus, the $H$ boson should be a
relatively narrow resonance, with $\Gamma(H\to all) = 30\,(5)$~GeV, decaying
predominantly to $t \bar t$ quarks, with a branching ratio
$B(H\to t \bar t )\sim 0.77\,(4)$. As for the other dominant decays,  
we estimate $B(H\to WW)\sim 0.11\,(2)$ and
$B(H\to ZZ)\sim B(H \to hh) \sim 0.05\,(1)$. Finally, the $H$ would mainly be
produced via gluon-gluon fusion (ggF), with the same SM cross section
$\sigma^{\rm ggF} (pp\to H) \sim 1000\,(200)$~fb \cite{yellow}.
From these values, we find peak cross sections in the various channels through
intermediate $ZZ$, $WW$, and $hh$ states, e.g.\ 
$\sigma_P(pp\to H  \to 4l) = 0.23\,(4)$~fb,
$\sigma_P(pp\to H\to 2l2\nu)= 5\,(1)$~fb,
$\sigma_P(pp\to H \to b\bar b + \gamma\gamma)= 0.13\,(3)$~fb and so on.
These values are comparable to (or smaller than) the corresponding background
contributions $\sigma_B$. Hence, we stress the importance of properly taking
into account the interference term, proportional to
$\pm\sqrt{\sigma_P\sigma_B}$, with its characteristic change of sign around
the peak. 

In the following we will recapitulate the experimental indications that
support the existence of a new resonance in the expected mass range. In some
cases, where numerical data are available, we will also report the results
from fits that include a resonance. 

\subsection{The ATLAS \boldmath{$t \bar t$} events}
\begin{figure}[!b]
\centering
\includegraphics[width=0.25\textwidth,clip]{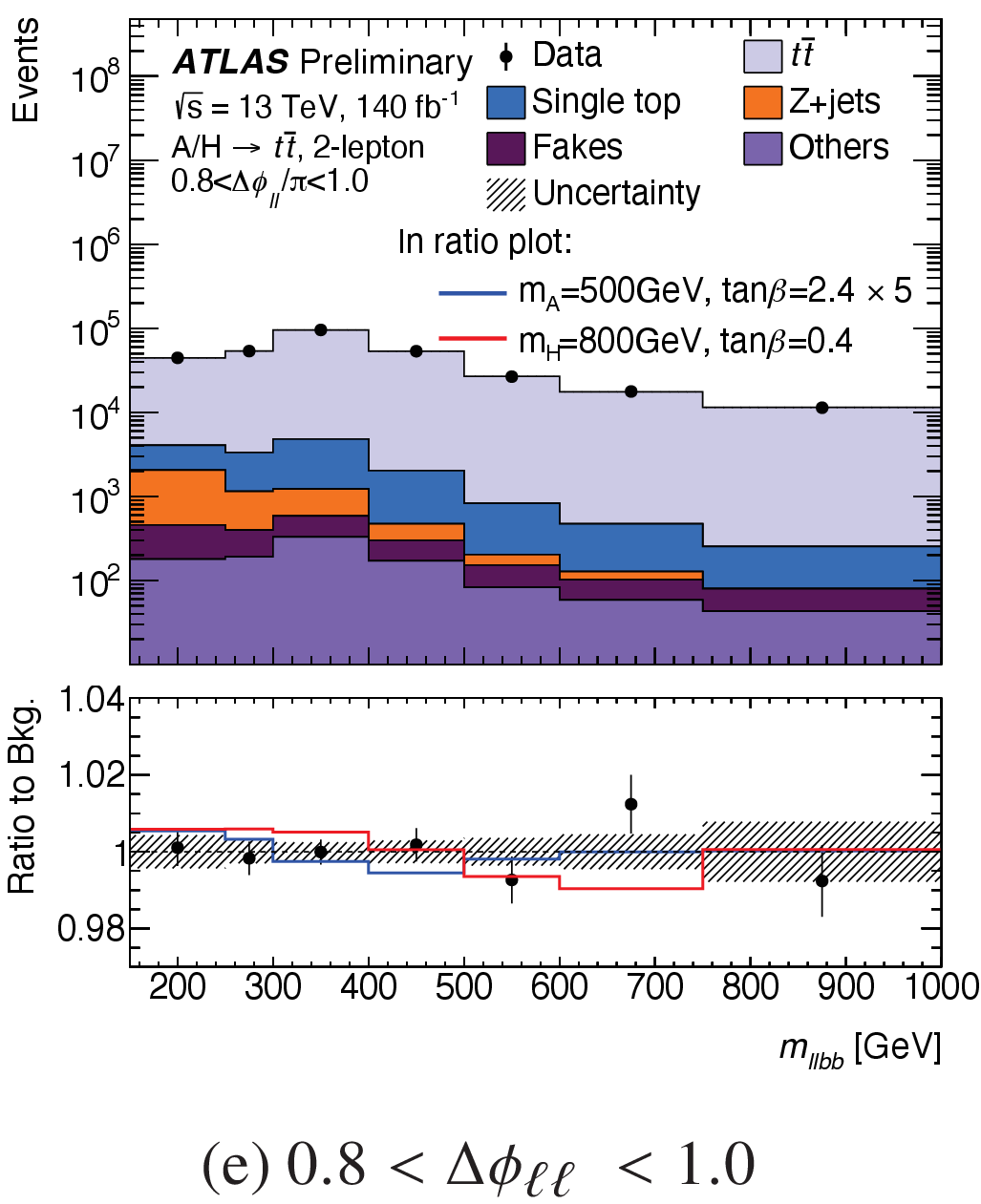}
\caption{\it ATLAS slight excess of $t \bar t$ pairs 
for an invariant mass of the $llbb$ system around 675~GeV. }
\label{ttbar}
\end{figure}

Owing to the large branching ratio, the first place to look for 
the new resonance $H$ is the $t\bar t$ channel. However, in
the relevant energy region $m(t\bar t)=620\div820$~GeV the background cross
section is about 100 times larger \cite{CMS_top} than the expected signal
$\sigma(pp\to H \to t \bar t) \lesssim 1$~pb. Nevertheless, a slight excess
was observed by ATLAS \cite{ATLAS_conf} in our mass region  around~675 GeV;
see Fig.~\ref{ttbar}. It is just a $1\%$ excess, but this is precisely the
expected value. Therefore, it represents an interesting indication, even
though we are speaking of a $1\sigma$ effect.

\subsection{The ATLAS 4-lepton events}
\begin{figure}[hb]
\centering
\includegraphics[width=0.40\textwidth,clip]{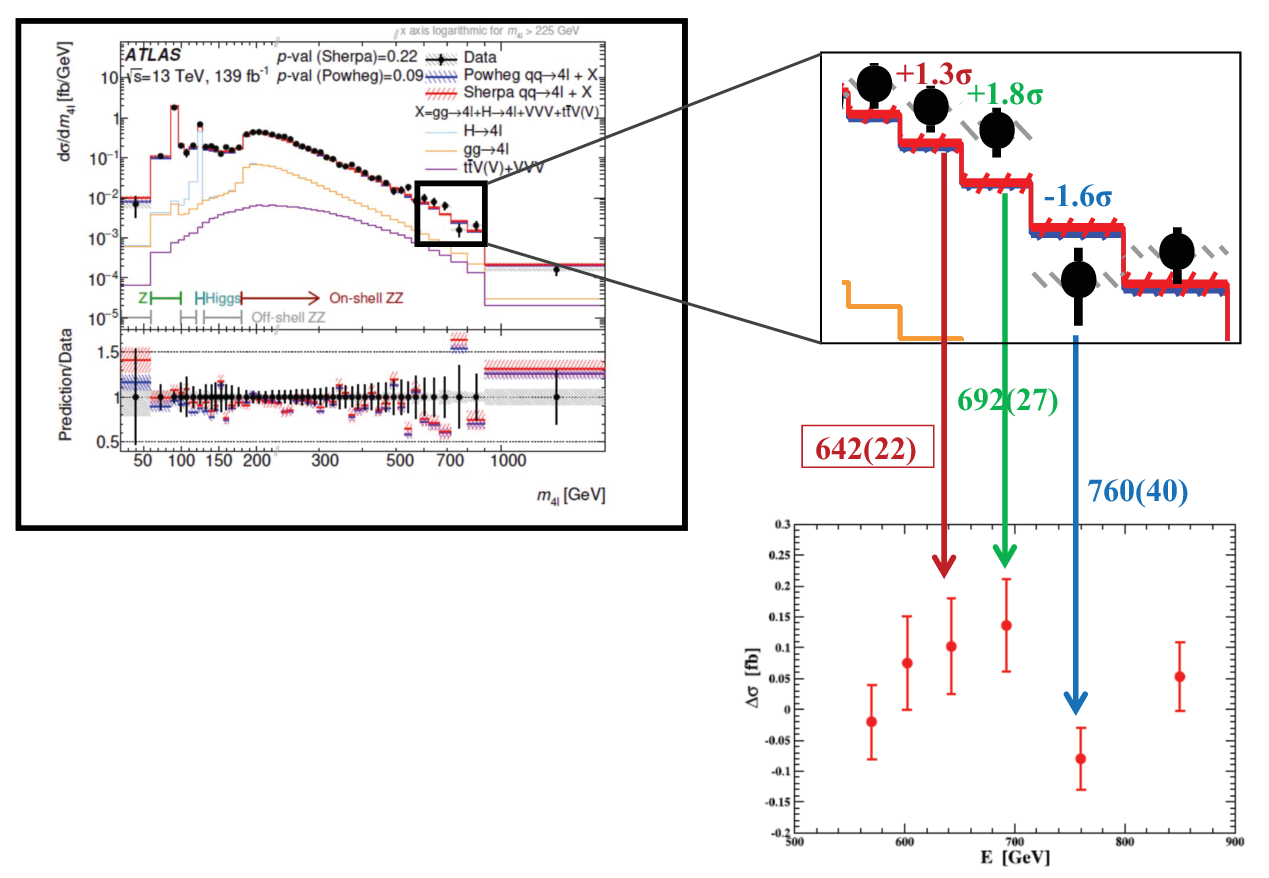}
\caption{\it The ATLAS four-lepton differential cross section
\cite{atlasnew} and the difference
$\Delta \sigma=\sigma_{\rm EXP} - \rm {\sigma}_{\rm B} $ in
Table~\ref{leptonxsection}.}
\label{ATLAS_cross_section}
\end{figure}

\begin{figure}[t]
\centering
\includegraphics[width=0.30\textwidth,clip]{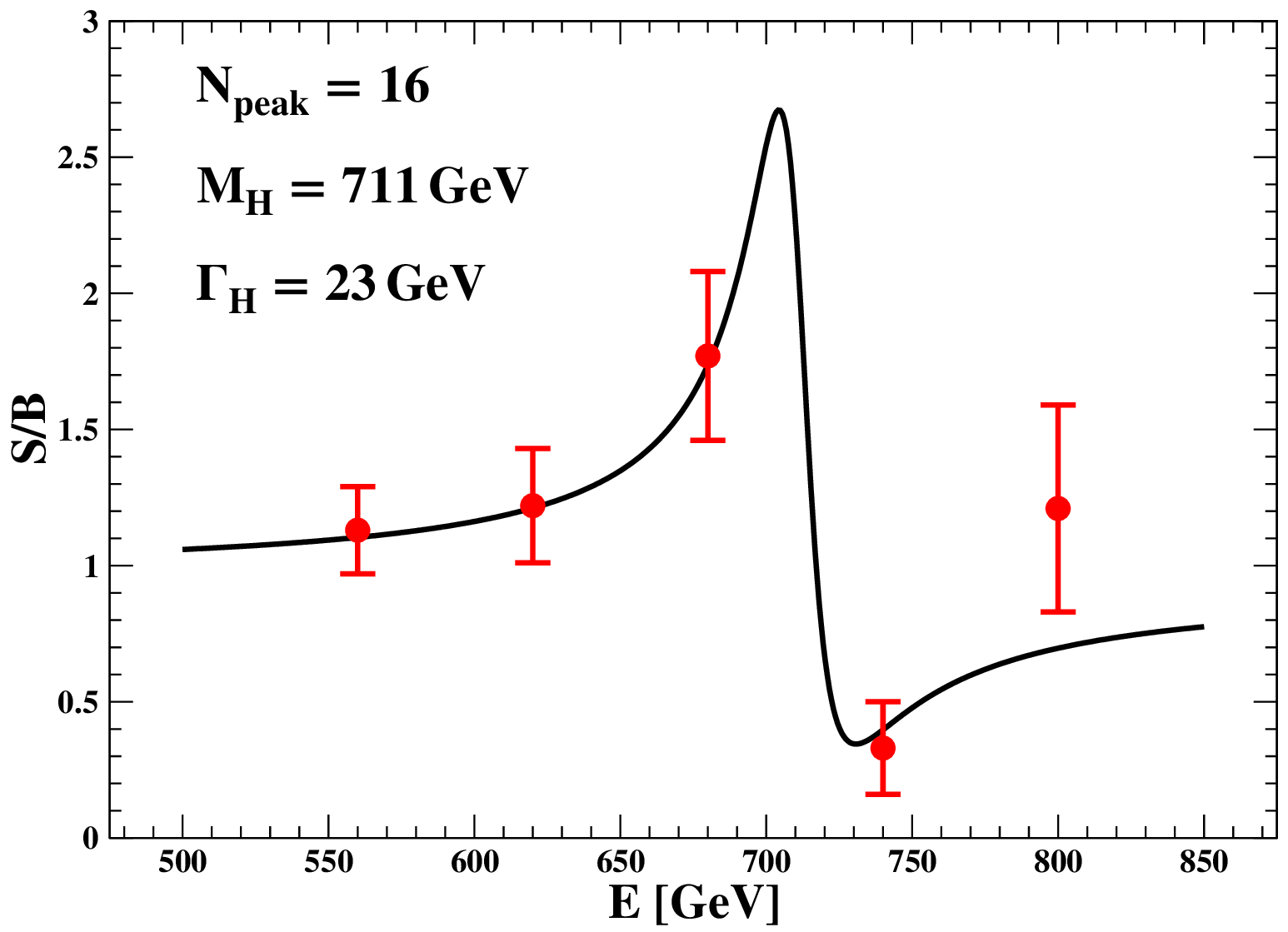}
\caption{\it A fit to the data in Table~\ref{ATLAS-MVA} yields
$M_H= 711\,(24)$~GeV, $\Gamma_H= 23\,(10)$~GeV and $\sigma_P= 0.30\,(18)$~fb.
For the given acceptances and luminosity, this gives a resonant peak of events
$N_{\rm peak}= 16\,(10)$.}
\label{fit2giugno}
\end{figure}

Next to consider is the ATLAS charged four-lepton final state.  
The average differential cross section $\langle d \sigma/d E\rangle$
\cite{atlasnew}, with $E=m(4l)$,  is reported in the upper panel 
of Fig.~\ref{ATLAS_cross_section}. From these average values, we may thus
determine the integrated cross section in the various bins. The numerical
values are presented in Table~\ref{leptonxsection} and the difference
$\Delta \sigma = {\sigma}_{\rm EXP} - { \sigma}_{\rm B} $  is shown in the
lower panel of Fig.~\ref{ATLAS_cross_section}. These data indicate an
excess/defect sequence of the type that could be produced by the change of
sign of the interference $(M^2 -s )$ across a resonance peak. In
Ref.~\cite{EPJC} a fit to these data was performed, by first separating out
the non-ggF background from the pure $gg \to 4l$ contribution, and then using
the latter part to compute the resonance-background interference. While
referring to Tables~5 and 6 of Ref.~\cite{EPJC}, here we just give the results
of a fit to the subtracted data, namely $M_H= 677^{+30}_{-14}$~GeV,
$\Gamma_H= 21^{+28}_{-16}$~GeV, and $\sigma_P= 0.40^{+0.62}_{-0.34}$~fb,
consistently with our expectations. The same``peak/dip'' pattern is found in
another ATLAS analysis \cite{atlas4lHEPData}, where the charged four-lepton
events were classified according to their topology. The dominant ggF-like
events, depending on their contamination with the background, are then
separated in four mutually exclusive categories. Their values and the
resulting $S/B$ are reported in Table~\ref{ATLAS-MVA} (for simplicity we
denote by $S/B$ the ratio of experimental to background cross sections or
of observed to expected events). Again, the results of a fit are consistent
with our expectations; see Fig.~\ref{fit2giugno}.
\begin{table*}[t]
\caption{\it The ATLAS \cite{atlasnew} four-lepton cross section and
background for $m(4l) = 555 \div 900$~GeV.}.
\begin{center}
\begin{tabular}{cccc}
Bin [GeV]& ${\sigma}_{\rm EXP} $~[fb]  & $\rm{ \sigma}_{\rm B} $~[fb] &
$(\sigma_{\rm EXP} - \rm {\sigma}_{\rm B} $)~[fb]  \\
\hline \hline
555--585 & 0.252 $\pm 0.056$ & $0.272 \pm 0.023$ & $-0.020 \pm 0.060$  \\
\hline
585--620 & $0.344\pm 0.070$ & $ 0.259 \pm 0.021 $ & $+0.085 \pm 0.075$ \\
\hline
620--665 & $0.356 \pm 0.075$ & $ 0.254 \pm 0.023$ & $+0.102\pm 0.078$  \\
\hline
665--720 & $0.350 \pm 0.073$ &$ 0.214 \pm 0.019$  &$+0.136 \pm 0.075$ \\
\hline
720--800 & $0.126 \pm 0.047$ &$ 0.206 \pm 0.018$ & $ -0.080 \pm 0.050$ \\
\hline
800--900 & $0.205\pm 0.052$ &$ 0.152 \pm 0.017$  & $+0.053 \pm 0.055$  \\
\hline
\end{tabular}
\end{center}
\label{leptonxsection}
\end{table*}
\begin{table*}[t]
\caption{\it ATLAS \cite{atlas4lHEPData} various types of ggF-like 4-lepton events with, in parentheses, the expected background.}
\begin{center}
\begin{tabular}{ccccccc}
$\rm E$ [GeV] & high-4$\mu$ & high-2e2$\mu$ & high-4e & low &Total & S/B  \\
\hline
560\,(30) & 4 (3.6) & 3 (6.2) & 5 (2.7)  & 38 (32.0) &50 (44.5)&$1.13\pm0.16$\\
\hline
620\,(30) & 3 (2.3) & 2 (3.9) & 4 (1.7) & 25 (20.0)  &34 (27.9)&$1.22\pm0.21$\\
\hline
680\,(30) & 1 (1.5) & 2 (2.5) & 3 (1.1) & 26 (13.0)  &32 (18.1)&$1.77\pm0.31$\\
\hline
740\,(30) & 0 (1.0) & 1 (1.6) & 0 (0.7) &  3 (8.7) & 4 (12.0) & $0.33\pm0.17$\\
\hline
800\,(30) & 1 (0.7) & 2 (1.1) & 0 (0.5) & 7 (6.0) & 10 (8.3) & $1.21\pm0.38$\\
\hline
\end{tabular}
\end{center}
\label{ATLAS-MVA}
\end{table*}

\subsection{The CMS four-lepton events}
\begin{figure}[!ht]
\centering
\includegraphics[width=0.38\textwidth,clip]{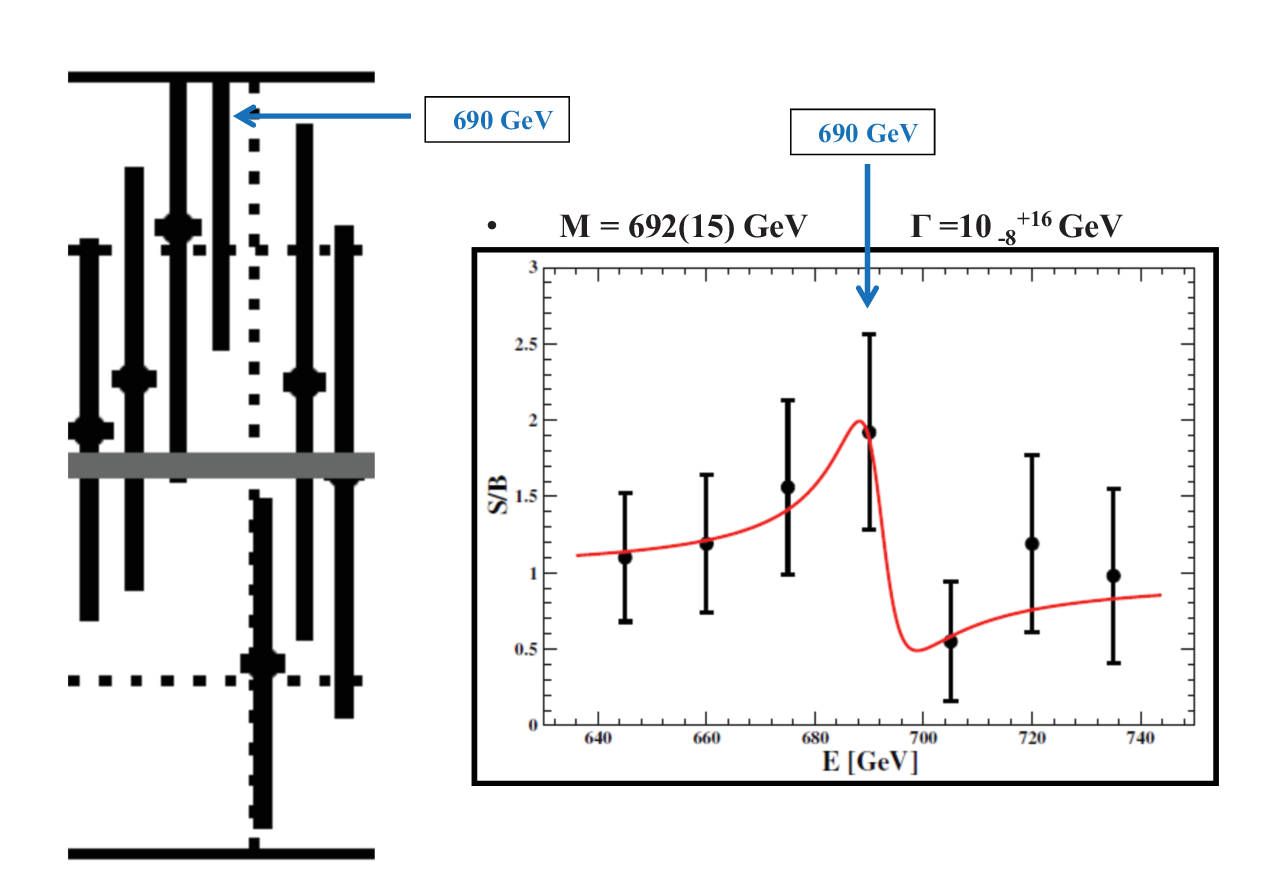}
\caption{\it The CMS four-lepton data \cite{CMS_4leptons_2024} for the $S/B$
ratio from 640 to 740~GeV and a fit with a resonance.}.
\label{CMS_7data}
\end{figure}

Figure~\ref{CMS_7data} shows (left) the CMS four-lepton data for the $S/B$
ratio \cite{CMS_4leptons_2024} and (right) a fit with a resonance. Note that
the peak is exactly at 690~GeV. Besides, the presence of a resonance could
also produce a defect of events due the change of sign of the interference
across the peak. In this case, the overall local deviation of the three points
at 675, 690, and 705 GeV is above $2\sigma$. 

\subsection{The ATLAS high-mass $\gamma\gamma$ events}
\begin{figure}[t]
\includegraphics[trim = 0mm 10mm 0mm 5mm,clip,width=6.0cm]
{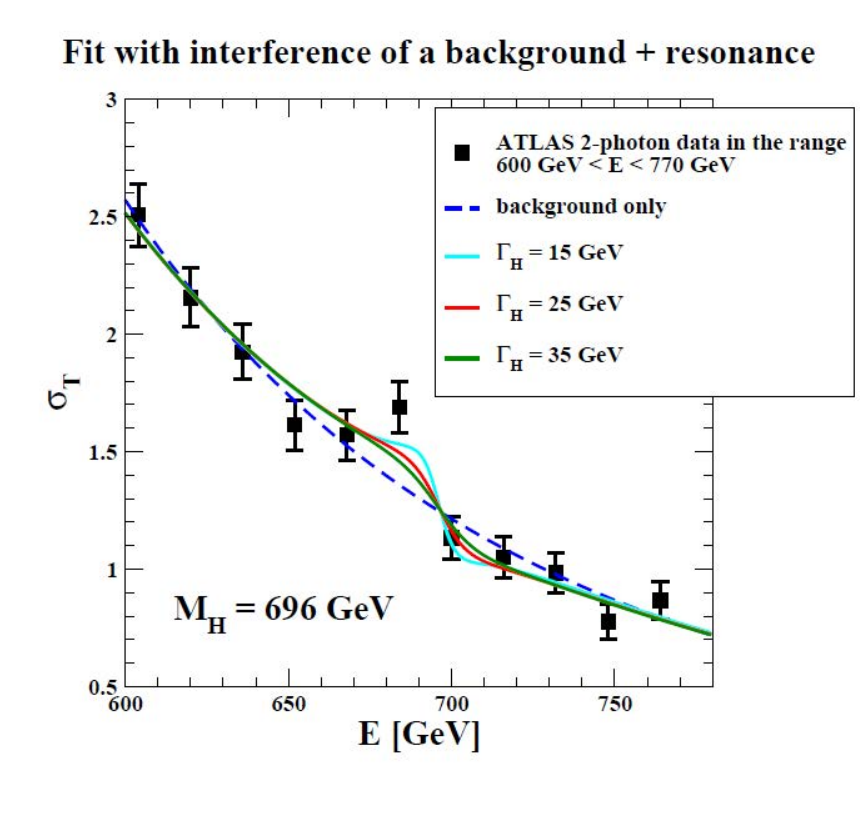}
\caption{\it Three fits to the ATLAS  $\gamma\gamma$ 
data reported in Fig.~3 of Ref.~\cite{atlas2gammaplb}.}
\label{twogamma1}
\end{figure}

In Ref.~\cite{EPJC} we also considered the ATLAS diphoton events
\cite{atlas2gammaplb} in the range $m(\gamma\gamma)=600\div770$~GeV. Here,
the background gives a good description of all points with the exception of a
sizeable excess at 684~GeV, estimated by ATLAS to have a local significance
of more than $3\sigma$. This shows how a relatively narrow resonance
might remain hidden behind a large background nearly everywhere, the only
observable effect being a sharp interference effect. Referring to
Ref.~\cite{EPJC} for more details, we just report in Fig.~\ref{twogamma1} a
typical fit to the data. 

\subsection{The ATLAS and CMS $b \bar b+ \gamma\gamma$ channel}
\begin{figure}[h!]
\centering
\includegraphics[width=0.35\textwidth,clip]{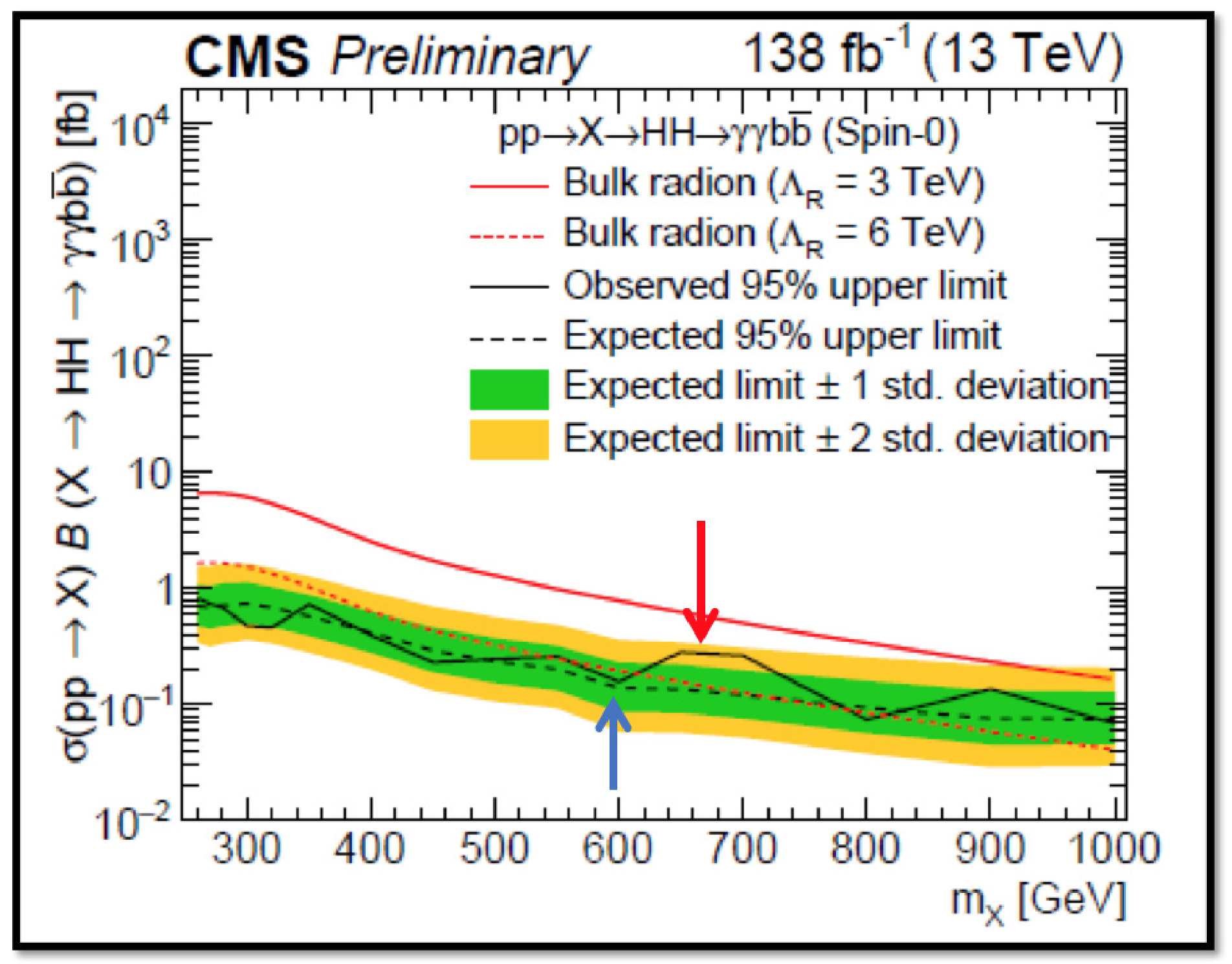}
\caption{\it CMS observed and expected 95$\%$ upper limits for
$\sigma (pp\to X) B(X\to hh \to b \bar b + \gamma\gamma)$ 
\cite{CMS_BBGG_paper}.}
\label{CMS_DBGG}
\end{figure}
The ATLAS and CMS Collaborations also searched for new resonances decaying,
through a pair of $h=h(125)$ scalars, into a $b\bar b$ quark pair and a
$\gamma\gamma$ pair. In particular, Ref.~\cite{CMS_BBGG_paper} gives
a 95$\%$ upper limit for the cross section of the full process:
\BE
\sigma({\rm full})\; = \;\sigma(pp\to X) B(X\to hh\to b\bar b+\gamma\gamma)\;.
\EE
In the plateau 650$\div$700~GeV, the observed limit was
$\sigma^{\rm obs}({\rm full}) \sim 0.27$~fb, to be compared with the expected
value (with $1\sigma$ uncertainty)
$\sigma^{\rm expected}({\rm full})\sim 0.13(8)$~fb; see Fig.~\ref{CMS_DBGG}.
The statistical significance of the excess is modest, about $1.6\sigma$, but
the energy region and the size of the excess fit well with our expectations
for the second resonance. Indeed, the observed excess
$\Delta \sigma({\rm full}) = +0.14\,(8)$~fb, if interpreted in terms of the
$H$ resonance, would agree well with our previous estimate
$\sigma^{\rm Theor}_P(pp\to H\to hh\to  b\bar b+\gamma\gamma)=0.13\,(3)$~fb. 
\begin{figure}[h!]
\centering
\includegraphics[width=0.38\textwidth,clip]{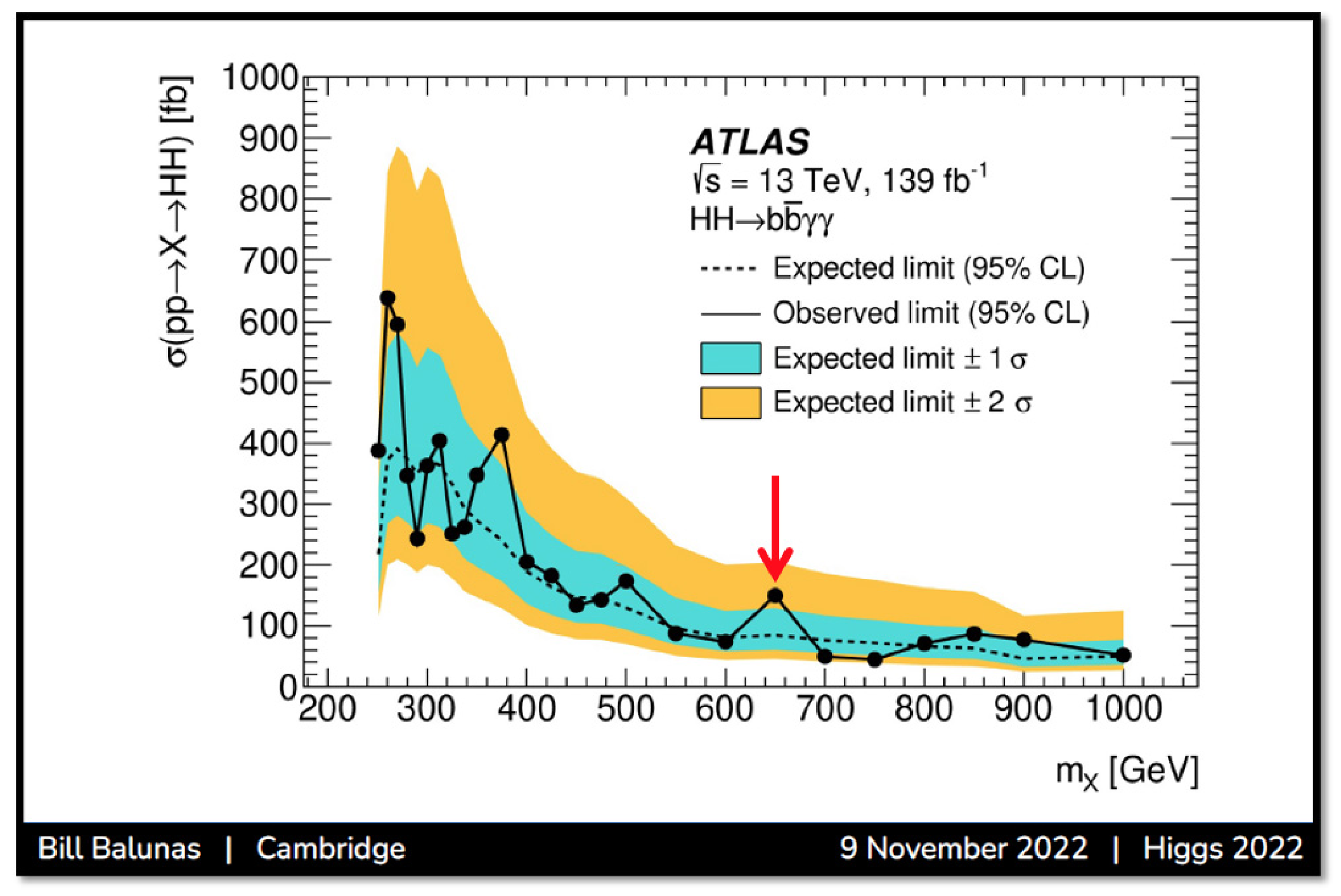}
\caption{\it ATLAS 95$\%$ upper limits for $\sigma (pp\to X \to h(125)h(125))$ 
from the $(b \bar b + \gamma\gamma)$ channel \cite{ATLAS_BBGG_paper}.}
\label{ATLAS_BBGG}
\end{figure}
The analogous ATLAS plot is shown in Fig.~\ref{ATLAS_BBGG} and the numerical
values are given in Table~\ref{sigmas}. Note that the sizes of the bins at
700~GeV are not the same for ATLAS and CMS data, while ATLAS also has a bin
centred at 750~GeV. This is why, apart from 650~GeV, the results at that
higher energies cannot be easily compared. Nevertheless, the excess observed
at 650~GeV, namely $\Delta \sigma(pp\to H\to hh)= +65^{+24}_{-44}$~fb with
$1\sigma$ errors, corresponding to
$\Delta\sigma({\rm full})=+0.17^{+0.06}_{-0.12}$~fb, fits well with the CMS
value at the same energy and with our expectation of a peak value
$ \sigma^{\rm Theor}_P(pp\to H\to hh) \sim 50\,(10)$~fb. 
\begin{table}[b!]
\caption{\it The ATLAS 95$\%$ observed upper limits $\sigma^{\rm obs}$ on the
cross section $\sigma(pp\to X \to h h)$, extracted from the
$b\bar b + \gamma\gamma$ final state \cite{ATLAS_BBGG_paper}. In the third
column, we report the corresponding upper limits on the expected background,
with $\pm 1\sigma$ and $\pm 2\sigma$ theoretical uncertainties (see the
HEPData file of Ref.~\cite{ATLAS_BBGG_paper}.)}
\begin{center}
\begin{tabular}{ccc}
\hline\hline
$M_X$ (GeV) & $\sigma^{\rm obs}$ [fb] & $\sigma^{\rm expected}$ [fb] \\ \hline
600 & 73.6 & $81.1^{ {+43.3}^{+119.0} }_{ {-22.7}_{-37.6} }$ \\ \hline
650 & 149.3 & $84.4^{ {+44.4}^{+120.1} }_{ {-23.6}_{-39.1} }$ \\ \hline
700 & 49.4 & $76.5^{ {+40.0}^{+109.6} }_{ {-21.4}_{-35.4} }$ \\ \hline
750 & 44.5 & $71.7^{ {+37.6}^{+103.3} }_{ {-20.0}_{-33.2} }$ \\ \hline
800 & 71.0 &  $65.8^{ {+35.1}^{+96.5} }_{ {-18.4}_{-30.5} }$ \\ \hline
\end{tabular}
\end{center}
\label{sigmas}
\end{table}

The first impression from these ATLAS measurements is that of a modest
$+1.2\sigma$ excess at 650~GeV, followed by two slight $-1.3\sigma$ defects.
However, this is a very conservative estimate. In fact, as already emphasised
in Ref.~\cite{EPL}, in our mass region and to a very good approximation, the
theoretical uncertainties simply shift the central values up and down.
Therefore, if we compute the difference between the observed values at
650 and 700~GeV, which is $(149.3-49.4)$~fb $=99.9$~fb, the result becomes much
larger than the corresponding differences between the central values
$(84.4-76.5)$~fb $=7.9$~fb or along the $+1\sigma$ and
$+2\sigma$ contours, viz.\ $(128.8-116.5)$~fb $=12.3$~fb and
$(204.5-186.1)$~fb $=18.4$~fb, respectively. Understanding this sizeable
discrepancy thus requires an {\it asymmetric} \/effect that increases the
number of events with $m(b \bar b + \gamma\gamma)$ around 650~GeV and, at the
same time, lowers the corresponding number with $m(b \bar b + \gamma\gamma)$
around 700~GeV and 750~GeV, like with a resonance around 675~GeV, with a decay
width of about 25~GeV and a peak cross section comparable to the background.
But apart from that, if we limit ourselves to the two excesses at 650 GeV,
the $+1.6\sigma$ of CMS and the $+1.2\sigma$ of ATLAS add up to a combined
effect of $2\sigma$.

\subsection{CMS-TOTEM $\gamma\gamma$ pairs produced in $pp$
diffractive scattering}
\begin{figure}[h!]
\centering
\includegraphics[trim = 0mm 15mm 0mm 15mm,clip,width=0.48\textwidth]
{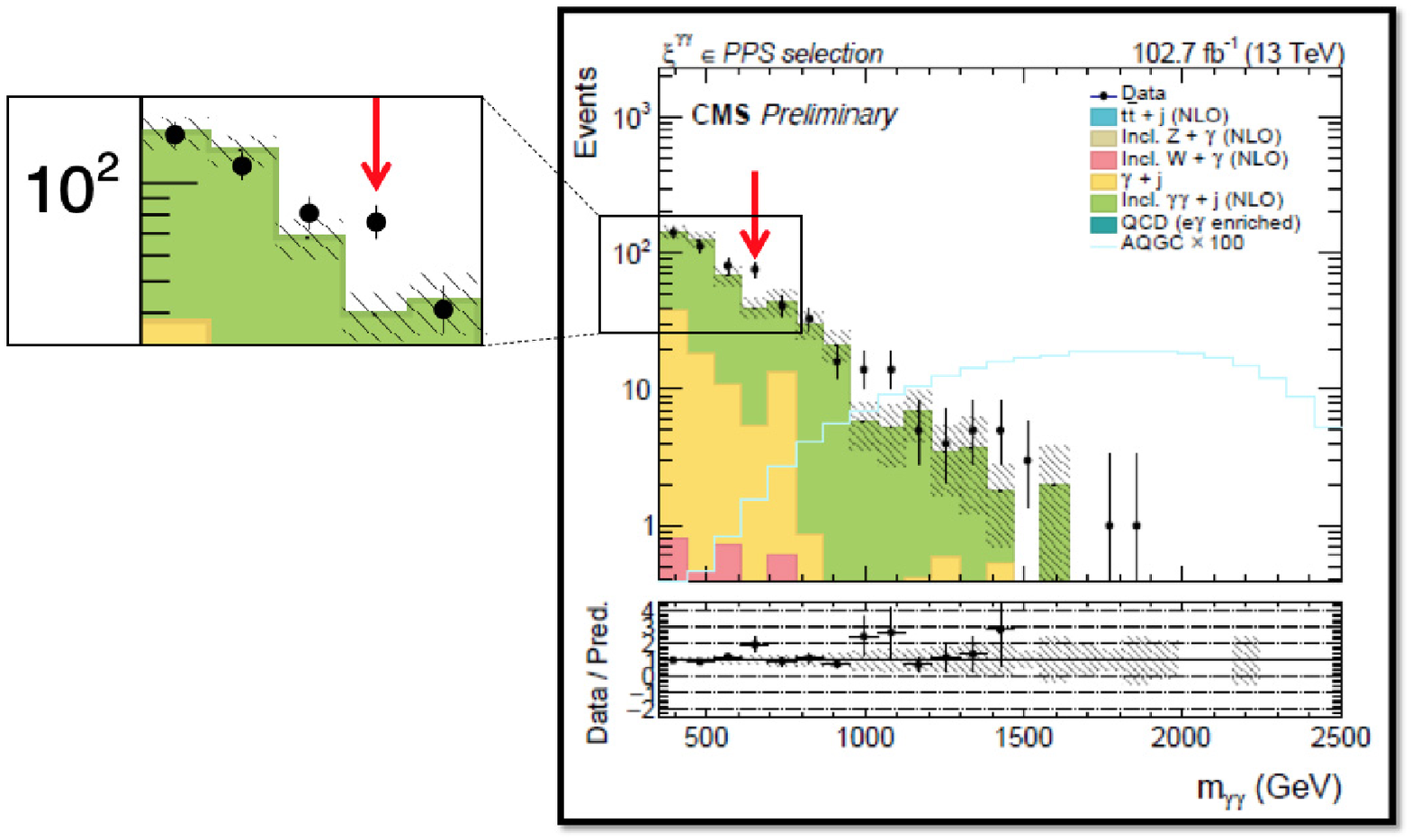}
\caption{\it The $\gamma\gamma$ events produced in $pp$ diffractive scattering
\cite{PRD_CMS_TOTEM}. In the range $650\,(40)$~GeV, the observed number of
events was $N_{\rm OBS}\sim 76\,(9)$, to be compared with an estimated
background $N_{\rm BKG}\sim 40\,(6)$.}
\label{diffractive}
\end{figure}
The CMS and TOTEM Collaborations have been searching for
high-mass photon pairs produced in $pp$ diffractive scattering, i.e., when
both final protons are tagged and have a large $x_F$. For our purpose, the
relevant information is contained in Fig.~\ref{diffractive} from
Ref.~\cite{PRD_CMS_TOTEM}. In the range of invariant mass $650\,(40)$~GeV and
for a statistics of 102.7~fb$^{-1}$, the observed number of $\gamma\gamma$
events was $N_{\rm obs}\sim76\,(9)$, to be compared with an estimated
background $N_{\rm B}\sim 40\,(9)$, which is quoted by CMS as being the best
estimate. In the most conservative case, i.e. $N_{\rm B}=49$, this is a local
$3\sigma$ effect and the only statistically significant excess in the plot.

\section{Summary and conclusions}

Motivated by the criticism in Ref.~\cite{cline}, we have summarised in the
present paper the essence of our previous work, which has largely been 
spread over Refs.~\cite{Cosmai2020,EPJC,EPL}. Thus, we have first recalled in
Sec.~2 the theoretical arguments supporting the existence of two mass
parameters instead of just one, viz.\ $m_h$ and $M_H$. These play different
roles in the Gaussian Effective Potential, because $m_h$ defines the quadratic
shape of the potential at its minimum, whereas the fourth power of $M_H$
determines the size of the zero-point energy and the potential depth. This
difference, which should be traced back to a description of SSB as a weak
first-order (or quasi-first-order) phase transition, can be intuitively
understood from Fig.~\ref{zpe}. We emphasise that the different scaling
$m^2_h \sim M^2_H L^{-1}$ of the two masses is crucial for the continuum
theory to have only one scale. However, in the cutoff theory we have both.
This is why the physical picture is substantially different from standard
perturbation theory, even though there is no contradiction with the technical
``triviality'' of the theory.

From the Gaussian Effective Action, these two masses are then seen to describe
the $p \to 0$ and larger-$p^2$ limits of the function $A(p,M_H)$ in
Eq.~(\ref{Ap}); for the ${\mathcal O}(N)$ theory we refer to Subsec.~2.5 of
Ref.~\cite{EPJC}. But a large Euclidean $p^2$ does {\it not} \/imply a large
$p^2_L$ in the Lorentzian signature. Therefore, the two regions, namely
$A(p,M_H)\sim \epsilon$ for $m_h$ and $A(p,M_H)\sim 1$ for $M_H$, indicate
two distinct mass-shell regions in Lorentzian energy-momentum space. The
existence of such a predicted two-mass structure has also been checked via
lattice simulations of the propagator, which lead to the definite prediction
$(M_H)^{\rm Theor} = 690\,(30)$~GeV. 

At the beginning of Sec.~3, we have also mentioned two preliminary points
before comparing with several experiments. First, due the large mass
$M_H\sim 690$~GeV in the scalar zero-point-energy, SSB would essentially be
induced within the pure scalar sector, as in the original SM perspective.
Secondly, the scalar self-coupling $\lambda(\mu)$ of the pure $\Phi^4$ theory
in Sec.~2 and the perturbative coupling $\lambda^{\rm (p)}(\mu)$, including
the effect of fermion and gauge fields, coincide at the Fermi scale
$v\sim 246$~GeV, where both take on the numerical value
$3 m^2_h/v^2\sim  0.77$. Therefore, with present experiments at the Fermi
scale, their large-$\mu$ difference remains unobservable. Confirming our
alternative description of SSB then requires observing the second resonance
and checking its phenomenology. 

From this point of view, the hypothetical $H$ differs from a conventional
Higgs boson of 700~GeV, because it would represent a relatively narrow
resonance with a decay width of $25 \div 35$~GeV. A phenomenological
comparison with LHC data was then presented in Subsecs.~$3.1 \div 3.6$,
where we mentioned the deviations supporting a new resonance in the expected
mass range:
\begin{enumerate}[label=\alph*)]
\item
ATLAS $t \bar t$ final state, $E = 675\,(75)$~GeV;
\item
ATLAS four-lepton channel, $E = 600\div 800$~GeV (see
Tables~\ref{leptonxsection} and \ref{ATLAS-MVA});
\item
CMS four-lepton channel, $E = 640\div740$~GeV;
\item
ATLAS data for $m(\gamma\gamma)$, $E = 684\,(16)$~GeV;
\item
ATLAS and CMS data for $X\to hh\to b\bar b+\gamma\gamma$, $E=650\,(25)$~GeV,
\item
CMS-TOTEM $\gamma\gamma$ events produced in $pp$ diffractive scattering,
$E = 650\,(40)$~GeV.
\end{enumerate}
These six indications have a combined statistical significance that is far
from negligible and could hardly be due to a statistical fluctuation.\footnote
{Local deviations should not be downgraded (or at least not {\it
fully}) by the look-elsewhere effect when there is a precise mass prediction,
as in our case.}
Note that all these experimental indications were already presented in
Refs.~\cite{EPJC,EPL} but not mentioned in Ref.~\cite{cline}, in which the
only counterargument was an ATLAS plot in the narrow-width approximation. As
we have explained above, the narrow-width approximation is inadequate when
dealing with a resonance whose decay width is comparable to (or larger than)
the experimental resolution of the final state. In this case, one must also
consider deviations that show up as defects of events. 
\begin{figure*}[!t]
\centering
\includegraphics[width=0.70\textwidth,clip]{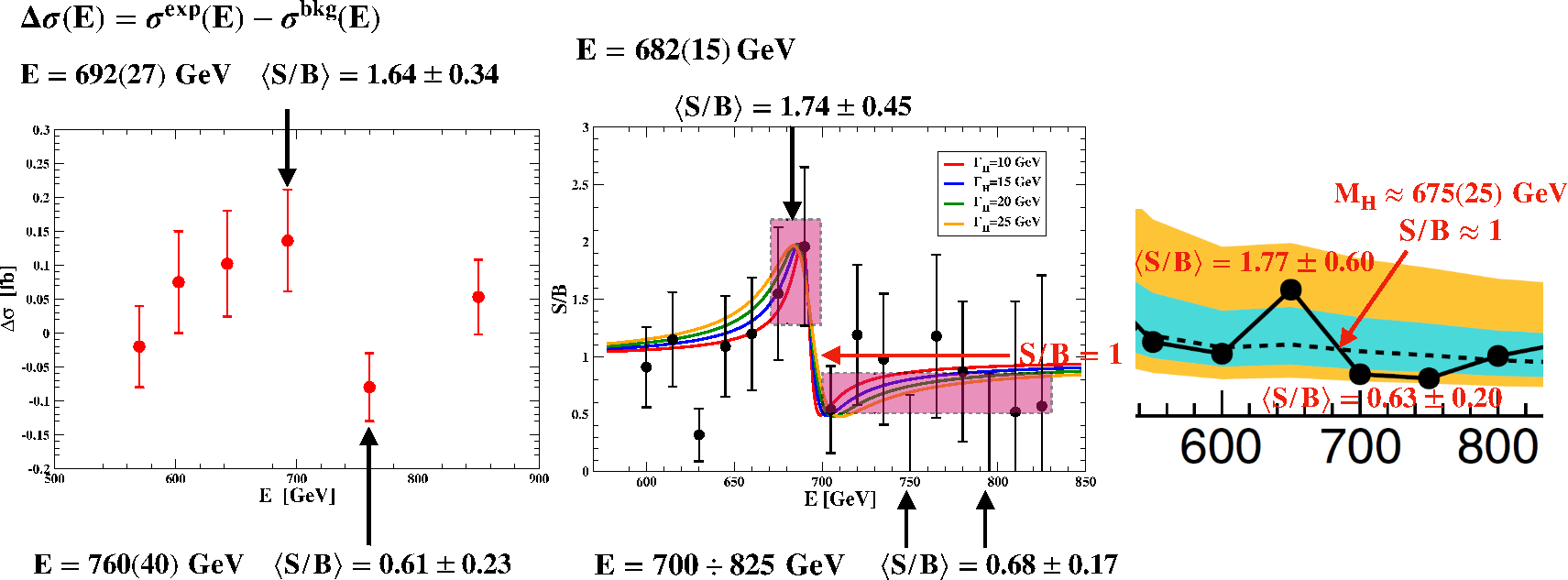}
\caption{The ATLAS four-lepton data of Fig.~\ref{ATLAS_cross_section},
together with the CMS data for $S/B$, from 600 to 825~GeV, and the analogous
ATLAS data \cite{ATLAS_BBGG_paper} for $\sigma(pp\to X\to hh)$ as extracted
from the $b\bar b + \gamma\gamma$ final state.}
\label{3-data}
\end{figure*}

To better appreciate this point, we put together in
Fig.~\ref{3-data} three sets of data: i) the ATLAS four-lepton cross sections
of Fig.~\ref{ATLAS_cross_section}; ii) the CMS $S/B$ ratio
(for $600\div 825$~GeV); iii) the ATLAS limits in Fig.~\ref{ATLAS_BBGG}. The
three sets indicate the same type of excess/defect pattern, with a combined
statistical significance which is far from negligible. Note the two empty bins
at 750 and 795~GeV in the CMS data, with their error bars representing the CMS
estimates for the upper limits that one might expect with more statistics,
i.e., $S/B<0.66$ and $S/B<0.86$, respectively. Together with the very low
point at 705~GeV, and in view of the large error bars of the remaining points,
this means that, in the region above 690~GeV, values with $S/B$ considerably
smaller than unity have a large probability content. The explicit check is
that the average ratio $\langle S/B\rangle^{\rm CMS}= 0.68\,(17)$, in the
range $705\div 825$~GeV, is in excellent agreement with the analogous
determination $\langle S/B\rangle^{\rm ATLAS}= 0.69\,(18)$ obtained from the
14 events observed by ATLAS, for $710\div830$~GeV, when compared to the
expected 20.3 events (see Table~\ref{ATLAS-MVA}). The theoretical curves,
especially those of green and yellow colour for widths $20\div25$~GeV, could
thus provide a clue with their prediction of a slow increase in $S/B$ from
about 0.5 at 700~GeV up to about 0.85 at 800~GeV. 
\begin{figure}[h!]
\centering
\includegraphics[width=0.50\textwidth,clip]{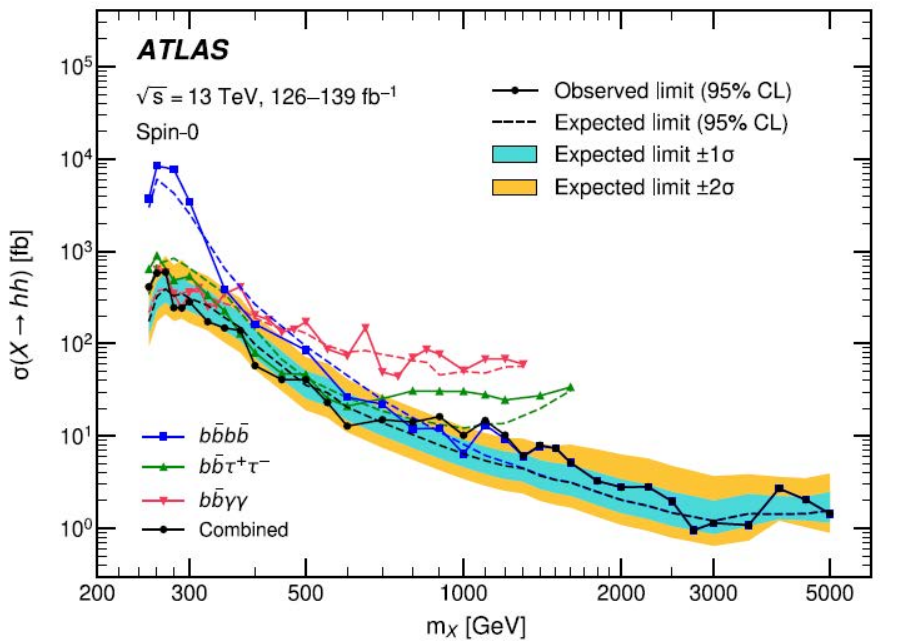}
\caption{\it The ATLAS limits on $\sigma(pp\to X \to hh)$ from
Ref.~\cite{ATLAS_DIHIGGS}. Note the different sizes of the energy bins:
50~GeV for $b \bar b \gamma\gamma$ (red) and 100~GeV for $b \bar b b \bar b$
as well as $b \bar b \tau^+\tau^-$, (blue and green) respectively. With such
larger bins, it is impossible to resolve the excess/defect pattern from the
interference $\pm  \sqrt { \sigma_P\sigma_B}$ with a relatively narrow
resonance, when the resonant peak $\sigma_P$ is comparable to the
background $\sigma_B$.}
\label{ATLAS_DIHIGGS}
\end{figure}

One may object that the LHC data we have collected are just a fraction of the
large number of ATLAS and CMS measurements. However, the final states we have
considered show a good resolution in invariant mass, differently from other
final states such as $b \bar b b \bar b$ or $b \bar b \tau^+\tau^-$. To
illustrate this point, let us look at Fig.~\ref{ATLAS_DIHIGGS} taken from the
ATLAS paper \cite{ATLAS_DIHIGGS}. In our region of interest, the limits from
$b \bar b \gamma\gamma$ in red are extracted from data in bins of 50~GeV,
while those from $b \bar b b \bar b$ or $b \bar b \tau^+\tau^-$, respectively
in blue and green, from bins of 100~GeV. As we have emphasised in Subsec.~3.5,
the relative values of the curve in red at 650, 700, and 750 GeV cannot be
understood in terms of the $1\sigma$ and $2\sigma$ uncertainties in the
background, because, in our region of interest, these uncertainties translate
rigidly. Therefore, one needs an asymmetric effect that increases the data at
650~GeV and lowers those at 700 and 750~GeV, as with a relatively narrow
resonance, with a mass of about 675~GeV and a width of about 25~GeV. This
could indeed produce the characteristic excess/defect pattern, in the range
$650\div750$~GeV, which is also seen in other channels with good energy
resolution; see Fig.~\ref{3-data}. Instead, with the larger bin size of the
blue and green data this is not possible. For instance, with such a bin size
and with respect to the background, one would just obtain a combined
$(65-54)=11$~fb from the $b \bar b \gamma\gamma$ channel, or a combined
$(14-8)=6$ number of extra four-lepton events in Table~3. This is why the
combined black line in Fig.~\ref{ATLAS_DIHIGGS} is not necessarily a more
faithful representation of the underlying physics. However, with the much
larger statistics available after including the RUN3 data, it should be
possible to reduce the bin size everywhere and clarify this point.


\begin{thebibliography}{24}

\bibitem{cline}
J.~M.~Cline,
``There is no 690 GeV resonance'',
{\it Nucl.\ Phys.\ B} \/\textbf{1029} (2026) 117526,
arXiv:2509.20115 [hep-ph].

\bibitem{Cosmai2020}
M.~Consoli and L.~Cosmai,
``The mass scales of the Higgs field'',
{\it Int.\ J.\ Mod.\ Phys.\ A} \/\textbf{35} (2020) 2050103,
arXiv:2006.15378 [hep-ph].

\bibitem{EPJC}
M.~Consoli and G.~Rupp,
``Second resonance of the Higgs field: motivations, experimental signals,
unitarity constraints'',
{\it Eur.\ Phys.\ J.\ C} \/\textbf{84} (2024) 951,
arXiv:2308.01429 [hep-ph].

\bibitem{EPL}
M.~Consoli, L.~Cosmai, F.~Fabbri, and G.~Rupp,
``The 690 GeV scalar resonance'',
{\it Europhys.\ Lett.} \textbf{152} (2025) 14002,
arXiv:2509.06479[hep-ph].

\bibitem{kurt}
K.~Symanzik,
``A field theory with computable large-momenta behavior'',
{\it Lett.\ Nuovo Cim.} \textbf{6S2} (1973) 77.

\bibitem{Romatschke}
P.~Romatschke,
``What if $\phi4$ theory in 4 dimensions is non-trivial in the continuum?'',
{\it Phys.\ Lett.\ B} \/\textbf{847} (2023) 138270,
arXiv:2305.05678 [hep-th].

\bibitem{huang}
K.~Huang,
``An Asymptotically free phi**4 theory'',
arXiv:hep-ph/9310235 [hep-ph].

\bibitem{paul1987}
P.~M.~Stevenson,
``Dimensional continuation and the two $(\lambda\phi^4)_4$ theories'',
{\it Z.\ Phys.\ C} \/\textbf{35} (1987) 467.

\bibitem{lundow2009critical}
P.~H.~Lundow and K.~Markstr{\"o}m,
``Critical behavior of the Ising model on the four-dimensional cubic lattice'',
{\it Phys.\ Rev.\ E} \/\textbf{80} (2009) 031104. 

\bibitem{Lundow:2010en}
P.~H.~Lundow and K.~Markstr{\"o}m,
``Non-vanishing boundary effects and quasi-first order phase transitions
  in high dimensional Ising models'',
{\it Nucl.\ Phys.\ B} \/\textbf{845} (2011) 120,
arXiv:1010.5958 [cond-mat.stat-mech].

\bibitem{akiyama2019phase}
S.~Akiyama, Y.~Kuramashi, T.~Yamashita, and Y.~Yoshimura,
``Phase transition of four-dimensional Ising model with higher-order
  tensor renormalization group'',
{\it Phys.\ Rev. D} \/\textbf{100} (2019) 054510,
arXiv:1906.06060 [hep-lat].

\bibitem{giapponesi2}
S.~Akiyama, Y.~Kuramashi, and Y.~Yoshimura,
``Phase transition of four-dimensional lattice $\phi4$ theory with tensor
  renormalization group'',
{\it Phys.\ Rev.\ D} \/\textbf{104} (2021) 034507,
arXiv:2101.06953 [hep-lat].

\bibitem{lang}
C.~B.~Lang,
``Computer Stochastics in scalar quantum field theory'',
{\it NATO Sci.\ Ser.\ C} \/\textbf{449} (1994) 133,
arXiv:hep-lat/9312004.

\bibitem{heller}
U.~M.~Heller,
{\it Nucl.\ Phys.\ B Proc.\ Suppl.} \textbf{34} (1994) 101,
arXiv:hep-lat/9311058.

\bibitem{barnes}
T.~Barnes and G.~I.~Ghandour,
``Variational Treatment of the Effective Potential and Renormalization
  in Fermi-Bose Interacting Field Theories'',
{\it Phys.\ Rev.\ D} \/\textbf{22} (1980) 924.

\bibitem{gaussian}
P.~M.~Stevenson,
``The Gaussian Effective Potential. 2. Lambda phi**4 Field Theory'',
{\it Phys.\ Rev.\ D} \/\textbf{32} (1985) 1389.

\bibitem{ciancitto}
M.~Consoli and A.~Ciancitto,
``Indications of the occurrence of spontaneous symmetry breaking in
  massless lambda Phi**4'',
{\it Nucl.\ Phys.\ B}, \/\textbf{254} (1985) 653.

\bibitem{Coleman:1973jx}
S.~R.~Coleman and E.~J.~Weinberg,
``Radiative Corrections as the Origin of Spontaneous Symmetry Breaking'',
{\it Phys.\ Rev.\ D} \/textbf{7} (1973) 1888.

\bibitem{alles}
P.~M.~Stevenson, B.~Alles, and R.~Tarrach,
``O(n) Symmetric Lambda phi**4 Theory: The Gaussian Effective Potential
  Approach'',
{\it Phys.\ Rev.\ D} \/\textbf{35} (1987) 2407. 

\bibitem{okopinska}
A.~Okopinska,
``Goldstone bosons in the Gaussian approximation'',
{\it Phys.\ Lett.\ B} \/\textbf{375} (1996) 213, 
arXiv:hep-th/9508087.

\bibitem{donoghue}
J.~F.~Donoghue,
``A Critique of the Asymptotic Safety Program'',
{\it Front.\ in Phys.} \textbf{8} (2020) 56,
arXiv:1911.02967 [hep-th].

\bibitem{GZK}
K.~Greisen,
``End to the cosmic ray spectrum?'',
{\it Phys.\ Rev.\ Lett.} \textbf{16} (1966) 748;
G.~T.~Zatsepin and V.~A.~Kuzmin,
``Upper limit of the spectrum of cosmic rays'',
{\it JETP Lett.} \textbf{4} (1966) 78.

\bibitem{Stevenson2005}
P.~M.~Stevenson,
``Comparison of conventional RG theory with lattice data for the
  4-D Ising model'',
{\it Nucl.\ Phys.\ B} \/\textbf{729} (2005) 542,
arXiv:hep-lat/0507038.

\bibitem{bagger}
J.~Bagger and C.~Schmidt,
``Equivalence Theorem Redux'',
{\it Phys.\ Rev.\ D} \/\textbf{41} (1990) 264.

\bibitem{atlascouplings}
G.~Aad \textit{et al.} [ATLAS COLLABORATION],
``Combination of Searches for Higgs Boson Pair Production in pp Collisions
  at s=13 TeV with the ATLAS Detector'',
{\it Phys.\ Rev.\ Lett.} \textbf{133} (2024) 101801,
arXiv:2406.09971 [hep-ex].

\bibitem{CMScouplings}
A.'Hayrapetyan  \textit{et al.} [CMS COLLABORATION],
``Constraints on the Higgs boson self-coupling from the combination of single
  and double Higgs boson production in proton-proton collisions at s=13TeV'',
{\it Phys.\ Lett.\ B} \/\textbf{861} (2025) 139210,
arXiv:2407.13554 [hep-ex].

\bibitem{yellow}
LHCPhysics Web,
``BSM Higgs production cross sections at $\sqrt{s}=13$ TeV
(update in CERN Report4 2016)'',
{\tt https://twiki.cern.ch/twiki/bin/view/
LHCPhysics/CERNYellowReportPageBSMAt13TeV} 

\bibitem{CMS_top}
A.~M.~Sirunyan {\it et al.} [CMS Collaboration],
``Measurements of $\mathrm{t\overline{t}}$ differential cross sections in
proton-proton collisions at $\sqrt{s}=$ 13 TeV using events containing two
leptons'',
{\it JHEP} \/\textbf{02} (2019) 149,
arXiv:1811.06625 [hep-ex].

\bibitem{ATLAS_conf}
ATLAS Collaboration,
``Search for heavy neutral Higgs bosons decaying to a
top quark pair in 140 fb$^{-1}$ of proton–proton
collision data at  $\sqrt{s}=13$~TeV,
ATLAS-CONF-2024-001, 
{\tt https://cds.cern.ch/record/
2891813/files/ATLAS-CONF-2024-001.pdf} 

\bibitem{atlasnew}
G.~Aad \textit{et al.} [ATLAS Collaboration],
``Measurements of differential cross-sections in four-lepton events
  in 13 TeV proton-proton collisions with the ATLAS detector'',
{\it JHEP} \/\textbf{07} (2021) 005,
arXiv:2103.01918 [hep-ex].

\bibitem{atlas4lHEPData}
G.~Aad \textit{et al.} [ATLAS Collaboration],
``Search for heavy resonances decaying into a pair of $Z$ bosons in the
$\ell ^+\ell ^-\ell '^+\ell '^-$ and $\ell ^+\ell ^-\nu {{\bar{\nu }}}$
final states using 139 $\mathrm {fb}^{-1}$ of proton{\textendash}proton
collisions at $\sqrt{s} = 13\,$TeV with the ATLAS detector'',
{\it Eur.\ Phys.\ J.\ C} \/\textbf{81} (2021) 332,
arXiv:2009.14791 [hep-ex].

\bibitem{CMS_4leptons_2024}
CMS Collaboration,
``Search for a new heavy scalar resonance decaying to a pair
of Z bosons in the four-lepton final state in proton-proton
collisions at $\sqrt{s}$=13TeV'',
arXiv:2605.26462v1 [hep-ex]. 

\bibitem{atlas2gammaplb}
G.~Aad \textit{et al.} [ATLAS Collaboration],
``Search for resonances decaying into photon pairs in 139 fb$^{-1}$ of
  $pp$ collisions at $\sqrt {s}$=13 TeV with the ATLAS detector'',
{\it Phys.\ Lett.\ B} \/\textbf{822} (2021) 136651,
arXiv:2102.13405 [hep-ex].

\bibitem{CMS_BBGG_paper}
A.~Tumasyan \textit{et al.} [CMS],
``Search for a new resonance decaying into two spin-0 bosons in a final
state with two photons and two bottom quarks in proton-proton collisions
at $ \sqrt{s}=13$~TeV'',
{\it JHEP} \/\textbf{05} (2024) 316,
arXiv:2310.01643 [hep-ex].

\bibitem{ATLAS_BBGG_paper}
G.~Aad \textit{et al.} [ATLAS Collaboration],
``Search for Higgs boson pair production in the two bottom quarks
 plus two photons final state in $pp$ collisions at $\sqrt{s}=13$ TeV
 with the ATLAS detector,''
{\it Phys.\ Rev.\ D} \/\textbf{106} (2022) 052001,
arXiv:2112.11876 [hep-ex].

\bibitem{PRD_CMS_TOTEM}
A.~Tumasyan \textit{et al.} [TOTEM and CMS Collaborations],
``Search for high-mass exclusive diphoton production with tagged protons
  in proton-proton collisions at s=13{\,}{\,}TeV'',
{\it Phys.\ Rev.\ D} \/\textbf{110} (2024) 012010,
arXiv:2311.02725 [hep-ex].

\bibitem{ATLAS_DIHIGGS}
G.~Aad \textit{et al.} [ATLAS Collaboration],
``Combination of Searches for Resonant Higgs Boson Pair Production Using
  pp Collisions at s=13{\,}{\,}TeV with the ATLAS Detector'',
{\it Phys. Rev. Lett.} \textbf{132} (2024) 231801,
arXiv:2311.15956 [hep-ex].

\end{thebibliography}
\end{document}